\documentclass[11pt]{article}
\DeclareMathAlphabet{\scr}{U}{rsfs}{m}{n}

\pdfoutput=1
\usepackage{latexsym}
\usepackage{epsfig}
\usepackage[mathscr]{eucal}
\usepackage{amsfonts}
\usepackage{amscd}
\usepackage{cite}
\usepackage{array}
\usepackage{bbold}
\usepackage{amssymb}
\usepackage{colordvi}
\usepackage[centertags]{amsmath}
\usepackage{enumerate}
\usepackage{graphicx}
\usepackage{booktabs}
\usepackage{theorem}
\usepackage[footnotesize]{caption}
\usepackage{soul}
\usepackage{url}
\usepackage[hidelinks]{hyperref}
\usepackage{mcite}
\usepackage{slashed}
\usepackage[dvipsnames]{xcolor}
\usepackage{bbm}
\usepackage[utf8]{inputenc}

\allowdisplaybreaks{}

\graphicspath{{analysis/}}
\newcommand{\newc}{\newcommand}
\newc{\be}{\begin{equation}}
\newc{\ee}{\end{equation}}
\newc{\bea}{\begin{eqnarray}}
\newc{\eea}{\end{eqnarray}}
\newc{\ol}{\overline}
\newc{\wt}{\widetilde}
\newc{\bs}{\boldsymbol}
\newc{\m}{\mathcal}
\newc{\la}{\langle}
\newc{\ra}{\rangle}

\newcommand{\beq}{\begin{eqnarray}}
\newcommand{\eeq}{\end{eqnarray}}
\newcommand{\bpmatrix}{\begin{pmatrix}}
\newcommand{\epmatrix}{\end{pmatrix}}
\newcommand{\ba}{\begin{array}}
\newcommand{\ea}{\end{array}}

\renewcommand{\ol}{\text{1l}}

\renewcommand{\eqref}[1]{Eq.~(\ref{#1})}

\newcommand{\bc}{\begin{center}}
\newcommand{\ec}{\end{center}}

\begin{document}

\title{
\vspace*{-3.7cm}
\phantom{h} \hfill\mbox{\small KA-TP-02-2020}\\[-1.1cm]
\phantom{h} \hfill\mbox{\small LU TP 20-18}
%\vspace*{0.7cm}
\\[1cm]
%\vspace{13mm}
\textbf{The Dark Phases of the N2HDM \\[4mm]}}

\date{}
\author{
Isabell Engeln$^{1,\,}$,
Pedro Ferreira$^{2,3\,}$\footnote{E-mail:
\texttt{pmmferreira@fc.ul.pt}} ,\\
M.~Margarete M\"uhlleitner$^{1\,}$\footnote{E-mail:
\texttt{margarete.muehlleitner@kit.edu}} ,
Rui Santos$^{2,3\,}$\footnote{E-mail:
  \texttt{rasantos@fc.ul.pt}} ,
Jonas Wittbrodt$^{4\,}$\footnote{E-mail: \texttt{jonas.wittbrodt@thep.lu.se}}
\\[5mm]
{\small\it
$^1$Institute for Theoretical Physics, Karlsruhe Institute of Technology,} \\
{\small\it 76128 Karlsruhe, Germany}\\[3mm]
{\small\it
$^2$Centro de F\'{\i}sica Te\'{o}rica e Computacional,
    Faculdade de Ci\^{e}ncias,} \\
{\small \it    Universidade de Lisboa, Campo Grande, Edif\'{\i}cio C8
  1749-016 Lisboa, Portugal} \\[3mm]
{\small\it
$^3$ISEL -
 Instituto Superior de Engenharia de Lisboa,} \\
{\small \it   Instituto Polit\'ecnico de Lisboa
 1959-007 Lisboa, Portugal} \\[3mm]
{\small\it
$^4$Department of Astronomy and Theoretical Physics,
Lund University,}\\
{\small\it Sölvegatan 14A, 223 62 Lund, Sweden}\\[3mm]
}
\maketitle

\begin{abstract}
We discuss the dark phases of the Next-to-2-Higgs Doublet model. The model is an extension of the Standard Model
with an extra doublet and an extra singlet that has four distinct CP-conserving phases, three of which provide dark matter candidates.
We discuss in detail the vacuum structure of the different phases and the issue of stability at tree-level of each phase.
Taking into account the most relevant experimental and theoretical constraints, we found that there are combinations
of measurements at the Large Hadron Collider that could single out a specific phase. The measurement of $h_{125} \to \gamma \gamma$
together with the discovery of a new scalar with specific rates to $\tau^+ \tau^-$ or $\gamma \gamma$ could exclude some phases
and point to a  specific phase.
\end{abstract}
\thispagestyle{empty}
\vfill
\newpage
\setcounter{page}{1}

%%%%%%%%%%%%%%%%%%%%%%%%%%%%%%%%%%%%%%%%%%%%%%%%%%%%%%%

\section{Introduction}

After the discovery of the Higgs boson~\cite{Aad:2012tfa, Chatrchyan:2012xdj} a large number of extensions of the Standard Model (SM) were
explored at the Large Hadron Collider (LHC) by searching both for new particles and for deviations in the Higgs couplings to the
remaining SM particles. However, not only are there  no direct hints
of new physics so far but all Higgs
  rates  are in very
good agreement with the SM predictions. Still, there is clear evidence of new physics, and in particular the existence of Dark Matter (DM)
which will be the subject of the particular extension of the SM to be discussed in this work.

The existence of DM manifests itself in gravitational effects to  baryon acoustic oscillations in the cosmic microwave background radiation~\cite{Bertone:2004pz},
which has shown that the relic abundance of DM in the Universe is about $26\%$~\cite{Ade:2015xua,Patrignani:2016xqp,Adam:2015rua}. Although there
is no indication about the nature of DM, it is clear that a
particle with a mass around the scale of
electroweak symmetry breaking and an interaction
cross section with the SM particles of the order of the weak force processes can account for the observed relic abundance as well as for structure formation.
These candidates are called Weakly Interacting Massive Particles (WIMPs).

When considering extensions of the SM with a DM candidate one needs to take into account all the presently available constraints. In order to have
an SM-like Higgs boson of 125~GeV and a scalar DM candidate, the simplest
extension of the SM is just the addition of a singlet field either
real or complex~\cite{Silveira:1985rk, McDonald:1993ex,Burgess:2000yq}.
The additional singlet is neutral  with
  respect to the SM gauge groups and DM is stabilised by a
symmetry. The next simplest extension that ensures $\rho=1$ at
 tree level is the popular
Inert Doublet Model (IDM)~\cite{Deshpande:1977rw, Ma:2006km, Barbieri:2006dq, LopezHonorez:2006gr}, a 2-Higgs Doublet Model where only one of the doublets acquires a
vacuum expectation value (VEV). The dark doublet (and the dark Higgs) is protected by a $\mathbb{Z}_2$ symmetry. The new dark sector contains two charged and two neutral fields,
the lightest of which is the dark matter candidate.

The Next-to-2-Higgs-Doublet Model (N2HDM)~\cite{Chen:2013jvg, Drozd:2014yla, Jiang:2016jea, Muhlleitner:2016mzt}, is an extension of the scalar sector of the SM by one doublet and one real singlet.
%PF In this particular version
In the particular version of doublet plus singlet extension that we will be studying,
two  $\mathbb{Z}_2$ symmetries are enforced. Depending on the pattern of symmetry breaking one ends up with a model with no dark matter candidates,
or a model with one or two dark matter particles. When unbroken, one of the $\mathbb{Z}_2$ symmetries stabilises the additional doublet, while the other stabilises the additional singlet. Therefore, the model
has four distinct phases: one with no DM, one with a singlet-like DM particle, one with a doublet-like DM candidate and finally one with two DM candidates. We call  the phase with singlet-like
DM phase~\cite{Drozd:2014yla} the Dark Singlet Phase (DSP), the doublet-like phase is called Dark Doublet Phase (DDP) and the SM-like phase with the two unbroken $\mathbb{Z}_2$ symmetries is designated Full Dark Matter Phase (FDP).

In this work, we compare the three N2HDM dark phases and wherever relevant we also include the Broken Phase (BP),
%PF
where the vacuum breaks both $\mathbb{Z}_2$ symmetries and there is
%PF
no dark matter
candidate. The DSP and DDP have additional scalar particles that mix with the CP-even
scalar from the SM doublet giving rise to new final states.
We will discuss how to phenomenologically distinguish these two phases.  The comparison between the three phases (and between each of them and the SM) can only
be  performed in the 125 GeV Higgs ($h_{125}$) decays and couplings to the remaining SM particles.
This is accomplished by studying the decay $h_{125} \to \gamma \gamma$ where
an extra loop of charged Higgs scalars --- either from the dark or from
the visible phases --- contributes.

The structure of the paper is as follows. We start by defining and describing the model and its phases in section~\ref{sec:model}.
 In section~\ref{sec:bil} we
study the coexistence of minima of different phases, and analyse the vacuum structure of the model.
In the following section~ \ref{sec:scans}
we present the experimental and theoretical constraints imposed on the model. In section~\ref{sec:LHC} we discuss how the different phases can be probed at the Large
Hadron Collider and add a brief discussion on future colliders. Finally we conclude in
section~ \ref{sec:conclusions}. The relations between the physical quantities at each phase and the input parameters of the model are shown in the appendices.

%%%%%%%%%%%%%%%%%%%%%%%%%%%%%%%%%%%%%%%%%%%%%%%%%%%%%%%
\section{The N2HDM \label{sec:model}}

The N2HDM \cite{Chen:2013jvg,Drozd:2014yla,Jiang:2016jea,Muhlleitner:2016mzt} is an extension of the SM,
where a complex  $SU(2)_{L}$ doublet $\Phi_2$ with hypercharge $Y=+1$
and a real $SU(2)_{L}$ singlet $\Phi_S$ with $Y=0$
are added to the SM field content.
In this work we will consider the most general renormalisable scalar potential invariant under two $\mathbb{Z}_{2}$ symmetries:
the first is
\begin{align}
\mathbb{Z}^{(1)}_{2}:\quad \Phi_{1}\rightarrow \Phi_{1},\quad \Phi_{2}\rightarrow -\Phi_{2},\quad \Phi_{S}\rightarrow \Phi_{S}\,,\label{eq:Z2_1}
\end{align}
while the second is
\begin{align}
\mathbb{Z}^{(2)}_{2}:\quad \Phi_{1}\rightarrow \Phi_{1},\quad \Phi_{2}\rightarrow \Phi_{2},\quad \Phi_{S}\rightarrow -\Phi_{S}\label{eq:Z2_2} \, .
\end{align}
Both symmetries are exact and --- if not spontaneously broken --- will give rise to DM candidates after electroweak symmetry breaking (EWSB). The potential reads
\begin{align}
V_{\text{Scalar}} =&\enspace m_{11}^{2} \Phi_{1}^{\dagger} \Phi_{1} + m_{22}^{2} \Phi_{2}^{\dagger} \Phi_{2}
+ \dfrac{\lambda_{1}}{2} \left(\Phi_{1}^{\dagger} \Phi_{1}\right)^{2}
+ \dfrac{\lambda_{2}}{2} \left(\Phi_{2}^{\dagger} \Phi_{2}\right)^{2}\notag\\
&+\enspace \lambda_{3} \Phi_{1}^{\dagger} \Phi_{1} \Phi_{2}^{\dagger} \Phi_{2}
+ \lambda_{4} \Phi_{1}^{\dagger} \Phi_{2} \Phi_{2}^{\dagger} \Phi_{1}
+ \dfrac{\lambda_{5}}{2} \left[\left(\Phi_{1}^{\dagger} \Phi_{2}\right)^{2} + \text{h.c.}\right]\label{eq:scalpot}\\
&+\enspace \dfrac{1}{2} m_{s}^{2}\Phi_{S}^{2} + \dfrac{\lambda_{6}}{8} \Phi_{S}^{4} + \dfrac{\lambda_{7}}{2} \Phi_{1}^{\dagger} \Phi_{1} \Phi_{S}^{2} + \dfrac{\lambda_{8}}{2} \Phi_{2}^{\dagger} \Phi_{2} \Phi_{S}^{2}\,,\notag
\end{align}
where all 11 free parameters of the Lagrangian,
\begin{equation}
  m_{11}^2\,,\enspace
  m_{22}^2\,,\enspace
  m_S^2\,,\enspace
  \lambda_{1-8}\,,\enspace\label{eq:lagrangepars}
\end{equation}
are real, or can be made to be so via a trivial rephasing of one of the doublets. Note that for the discrete symmetries  to be exact we introduce no soft breaking terms
in the potential. In particular, the term
$m_{12}^2 ( \Phi_{1}^\dagger \Phi_2 + h.c.)$ that would softly break the $Z^{(1)}$ symmetry is absent. This term is often used in many versions of the 2HDM and N2HDM to allow
for a decoupling limit, with the introduction of the new mass scale $m_{12}^2$.
After EWSB, the fields can be parametrised in terms of the charged complex fields $\phi_i^+$ $(i\in\{1,2\})$, the neutral CP-even fields $\rho_I$ $(I\in\{1,2,s\})$ and the neutral CP-odd fields $\eta_i$ as follows
\begin{align}
\Phi_1 = \begin{pmatrix} \phi_{1}^{+} \\
\dfrac{1}{\sqrt{2}}\left(v_{1} + \rho_{1} + i\,\eta_{1}\right)
\end{pmatrix},\quad
\Phi_2 = \begin{pmatrix} \phi_{2}^{+} \\
\dfrac{1}{\sqrt{2}}\left(v_{2} + \rho_{2} + i\,\eta_{2}\right)
\end{pmatrix}, \quad
\Phi_S = v_{s} + \rho_{s}\,.
\end{align}
Requiring the VEVs
\begin{align}
\left<\Phi_i\right>=\begin{pmatrix}
0 \\ \frac{v_i}{\sqrt{2}}
\end{pmatrix} \qquad\text{and}\qquad\left<\Phi_S\right>=v_s\,,
\label{eq:n2hdm_generalvacuum}
\end{align}
which break the $SU(2)_L\times U(1)_Y$ down to $U(1)_{EM}$, and possibly also the
symmetries,
%PF
to be solutions of the stationarity equations leads to the following three conditions,
%to be stationary points of the potential leads to three
%\textcolor{red}{\st{stationary}} \JWr{minimum}{stationarity} conditions
\begin{subequations}
\begin{alignat}{3}
\left<\dfrac{\mathrm{d}V}{\mathrm{d}v_{1}}\right> =&\, 0 \enspace&\Rightarrow&\enspace & - m_{11}^{2} &= \dfrac{1}{2}\,  \left(v_{1}^{2}\lambda_{1} + v_2^2 \left(\lambda_3+\lambda_4+\lambda_5\right) + v_s^2 \lambda_7\right)\label{eq:tadpole1},\\
\left<\dfrac{\mathrm{d}V}{\mathrm{d}v_{2}}\right> =&\, 0 \enspace&\Rightarrow&\enspace & - m_{22}^{2} &= \dfrac{1}{2}\,  \left(v_1^2 \left(\lambda_3+\lambda_4+\lambda_5\right) + v_{2}^{2}\lambda_{2}  + v_s^2 \lambda_8\right)\label{eq:tadpole2},\\
\left<\dfrac{\mathrm{d}V}{\mathrm{d}v_{s}}\right> =&\, 0 \enspace&\Rightarrow&\enspace &- m_s^2\,\, &= \dfrac{1}{2}\, \left(v_1^2\lambda_7 + v_2^2\lambda_8 + v_s^2\lambda_6\right).\label{eq:tadpoleS}
\end{alignat}
\end{subequations}
If we consider only minima that are CP-conserving and non-charge breaking, we can distinguish four cases:
\begin{itemize}
\item
\textbf{The Broken Phase (BP)}
--- In this phase both doublets and the singlet acquire VEVs and consequently both $\mathbb{Z}_2$ symmetries are spontaneously broken by EWSB.
There are no dark matter candidates, and the scalar particle spectrum consists of three CP-even, one CP-odd and two charged scalars.
This phase, with an extra soft breaking term for $\mathbb{Z}^{(1)}_{2}$, has been thoroughly studied in~\cite{Muhlleitner:2016mzt}.

\item
\textbf{The Dark Doublet Phase (DDP)} --- This is the case where only one of the doublets (either $\Phi_1$ or $\Phi_2$) and the singlet acquire VEVs.
This phase is the N2HDM equivalent to the Inert Doublet Model of the 2HDM \cite{Deshpande:1977rw, Ma:2006km, Barbieri:2006dq, LopezHonorez:2006gr}.
The $\mathbb{Z}^{(1)}_{2}$ symmetry is exactly preserved while
$\mathbb{Z}^{(2)}_{2}$ is spontaneously broken. There are four dark sector particles ---  two neutral and a pair of charged scalars --- and one extra CP-even scalar that mixes with the CP-even component from
the doublet which acquires a VEV.

\item
\textbf{The Dark Singlet Phase (DSP)} --- In this phase both doublets but not the singlet acquire VEVs. Hence, $\mathbb{Z}^{(2)}_{2}$
remains unbroken and the dark matter candidate has its origin in the singlet field. This phase is essentially a 2HDM plus a dark
real singlet~\cite{Silveira:1985rk, McDonald:1993ex, Burgess:2000yq}. The model has two CP-even,
one CP-odd and a pair of charged scalars in the visible sector plus a singlet-like DM particle.

\item
\textbf{The Fully Dark Phase (FDP)} --- Finally, we will consider a phase where only one doublet acquires a VEV.
This means that both $\mathbb{Z}_2$ symmetries remain unbroken and only one doublet couples to SM fields. Therefore,
this model contains just one SM-like Higgs boson with additional
couplings to dark particles. No new non-dark scalar is present and two distinct darkness quantum numbers are separately conserved.
\end{itemize}

We want the Lagrangian of the theory for all four phases to be exactly the same before EWSB. The kinetic terms are the same because they are only
determined by the  $SU(2)_L$ and $U(1)_Y$ quantum numbers. As for the Yukawa Lagrangian, the singlet field does not couple to the
fermions and we have to choose a Yukawa sector of type I, where only one doublet couples to the fermions
in order to be able to compare all four phases based
  on the same Lagrangian. The Yukawa Lagrangian takes the form,
\begin{align}
\mathcal{L}_{\text{Yukawa}} = -\bar{Q}^T_L Y_{U}\widetilde{\Phi}_f U_R -\bar{Q}^T_L Y_{D}\Phi_f D_R -\bar{L}^T_L Y_{L}\Phi_f E_R + \text{h.c.}\,,
\end{align}
where $\Phi_f$ is the doublet that couples to fermions, $Y$ are three-dimensional
Yukawa coupling matrices in flavour space, the left-handed fermions are grouped into the doublets
\begin{align}
Q_L = \begin{pmatrix}
U_L \\ D_L
\end{pmatrix} = \begin{pmatrix}
\left(u_L, c_L, t_L\right)^T\\
\left(d_L, s_L, b_L\right)^T
\end{pmatrix}, \qquad L_L = \begin{pmatrix}
N_L\\E_L
\end{pmatrix} = \begin{pmatrix}
\left(\nu_{e,L}, \nu_{\mu,L}, \nu_{\tau,L}\right)^T \\
\left(e_L, \mu_L, \tau_L\right)^T
\end{pmatrix},
\end{align}
and the right-handed fermion into the singlets
\begin{align}
U_R = \left(u_R, c_R, t_R\right)^T, \quad
D_R = \left(d_R, s_R, b_R\right)^T, \quad
E_R =\left(e_R, \mu_R, \tau_R\right)^T.
\end{align}
and $\widetilde{\Phi}_f$ stands for $\epsilon_{ij}\Phi^*_f$ , with $\epsilon_{ij}$ given by
\begin{align}
\epsilon_{ij} = \begin{pmatrix}
0 & 1 \\
-1 & 0
\end{pmatrix}.
\end{align}
We will now describe the four phases in detail.

%%%%%%%%%%%%%%%%%%%%%%%%%%%%%%%%%%%%%%%%%%%%%%%%%%%%%%%
\subsection{The Broken Phase (BP)}
In the broken phase, both the doublets and the singlet acquire VEVs that break both $\mathbb{Z}^{(1)}_2$ and $\mathbb{Z}^{(2)}_2$.
Since the model was discussed in great detail in~\cite{Muhlleitner:2016mzt},
we will just very briefly review the features of the model needed for this study.

The charged and pseudoscalar mass
matrices are diagonalised via the rotation matrix
\beq
R_\beta = \left( \begin{array}{cc} c_\beta & s_\beta \\ - s_\beta &
    c_\beta \end{array} \right) \;,
\eeq
with $t_\beta= \frac{v_2}{v_1}$. Here and from now on we use the abbreviations
$\sin x \equiv s_x$, $\cos x \equiv c_x$ and $\tan x \equiv t_x$.
This yields the massless charged and neutral would-be Goldstone
bosons $G^\pm$ and $G^0$, the charged Higgs mass
eigenstates $H^\pm$ and the pseudoscalar mass eigenstate $A$.
There are three CP-even gauge eigenstates  $(\rho_1, \rho_2, \rho_S)$,
two from the doublets and one from the singlet. The corresponding
mass eigenstates $H_1$, $H_2$
and $H_3$, are obtained via the orthogonal mixing matrix $R$ parametrised as
\beq
R =\left( \begin{array}{ccc}
c_{\alpha_1} c_{\alpha_2} & s_{\alpha_1} c_{\alpha_2} & s_{\alpha_2}\\
-(c_{\alpha_1} s_{\alpha_2} s_{\alpha_3} + s_{\alpha_1} c_{\alpha_3})
& c_{\alpha_1} c_{\alpha_3} - s_{\alpha_1} s_{\alpha_2} s_{\alpha_3}
& c_{\alpha_2} s_{\alpha_3} \\
- c_{\alpha_1} s_{\alpha_2} c_{\alpha_3} + s_{\alpha_1} s_{\alpha_3} &
-(c_{\alpha_1} s_{\alpha_3} + s_{\alpha_1} s_{\alpha_2} c_{\alpha_3})
& c_{\alpha_2}  c_{\alpha_3}
\end{array} \right)
\label{eq:mixingmatrix}
\eeq
in terms of the mixing angles $\alpha_1$ to $\alpha_3$, chosen to be in the range
\beq
- \frac{\pi}{2} \le \alpha_{1,2,3} < \frac{\pi}{2} \;.
\eeq
The matrix $R$ is defined is such a way that
\beq
\left( \begin{array}{c} H_1 \\ H_2 \\ H_3 \end{array} \right) = R
\left( \begin{array}{c} \rho_1 \\ \rho_2 \\ \rho_S \end{array} \right)
\eeq
diagonalises the scalar mass matrix $M_{\text{scalar}}^2$,
\beq
R M_{\text{scalar}}^2 R^T = \mbox{diag}(m_{H_1}^2,m_{H_2}^2,m_{H_3}^2) \;.
\eeq
We take, by convention,
\beq
m_{H_1} \le m_{H_2} \le m_{H_3} \;.
\eeq
In the broken phase, the 11 parameters of the N2HDM, \eqref{eq:lagrangepars}, are expressed through the input parameters
\beq
\alpha_1 \; , \quad \alpha_2 \; , \quad \alpha_3 \; , \quad t_\beta \;, \quad v \; ,
\quad v_S \; , \quad m_{H_{1,2,3}} \;, \quad m_A \;, \quad m_{H^\pm} . \label{eq:n2hdminputpars}
\eeq

%In appendix~\ref{app:lamrels} we provide expressions for the quartic
%couplings in terms of these input parameters.

The Higgs couplings $H_i$ ($i=1,2,3$) to the
massive gauge bosons $V\equiv W,Z$ are written as
\beq
i \, g_{\mu\nu} \, c(H_i VV) \, g_{H^{\text SM} VV} \;, \label{eq:gaugecoupdef}
\eeq
where $g_{H^{\text SM} VV}$ is the SM Higgs coupling to the massive
gauge bosons, and the coupling
modifiers $c(H_i VV)$ are presented in Table~\ref{tab:gaugecoupn2hdm}.
\begin{table}
\begin{center}
 \begin{tabular}{cc}
\toprule
\multicolumn{2}{c}{$c(H_i VV)$} \\
\midrule
$H_1$ & $c_{\alpha_2} c_{\beta-\alpha_1}$ \\
$H_2$ & $-c_{\beta-\alpha_1} s_{\alpha_2} s_{\alpha_3} + c_{\alpha_3}
s_{\beta-\alpha_1}$ \\
$H_3$ & $-c_{\alpha_3} c_{\beta-\alpha_1} s_{\alpha_2} - s_{\alpha_3}
s_{\beta-\alpha_1}$ \\
\bottomrule
\end{tabular}
 \caption{The effective couplings $c(H_i VV)$ of the neutral CP-even
   N2HDM Higgs bosons
   $H_i$ to the massive gauge bosons $V=W,Z$. \label{tab:gaugecoupn2hdm}}
\end{center}
\end{table}
As previously discussed the four phases of the N2HDM can only be compared
for the Yukawa Type I. The Yukawa Lagrangian reads
\beq
{\cal L}_Y = - \sum_{i=1}^3 \frac{m_f}{v} c(H_i ff) \,
\bar{\psi}_f \psi_f H_i
\label{eq:lyukn2hdm}
\eeq
where the effective coupling factors $c(H_i ff)$ are shown in Table \ref{tab:effyukn2hdm}.
\begin{table}
\begin{center}
  \begin{tabular}{cccc}
\multicolumn{4}{c}{Type I} \\ \toprule
$c(H_i ff)$ & $u$ & $d$ & $l$ \\ \midrule
$H_1$ & $(c_{\alpha_2} s_{\alpha_1} )/s_\beta$ & $(c_{\alpha_2}
s_{\alpha_1}) / s_\beta$ & $(c_{\alpha_2} s_{\alpha_1})/s_\beta$ \\
$H_2$ & $(c_{\alpha_1} c_{\alpha_3} - s_{\alpha_1} s_{\alpha_2}
s_{\alpha_3})/s_\beta$ & $(c_{\alpha_1} c_{\alpha_3}- s_{\alpha_1}
s_{\alpha_2} s_{\alpha_3})/s_\beta$ & $(c_{\alpha_1} c_{\alpha_3}-
s_{\alpha_1} s_{\alpha_2} s_{\alpha_3})/s_\beta$ \\
$H_3$ & $-(c_{\alpha_1} s_{\alpha_3} + c_{\alpha_3} s_{\alpha_1}
s_{\alpha_2} )/s_\beta$ & $-(c_{\alpha_1} s_{\alpha_3} + c_{\alpha_3}
s_{\alpha_1} s_{\alpha_2} )/s_\beta$ & $-(c_{\alpha_1} s_{\alpha_3} +
c_{\alpha_3} s_{\alpha_1} s_{\alpha_2}) /s_\beta$ \\ \bottomrule\\
\end{tabular}
\caption{The effective Yukawa couplings $c(H_i ff)$ of the N2HDM Higgs
   bosons $H_i$, as defined in Eq.~(\ref{eq:lyukn2hdm}) for Type I. \label{tab:effyukn2hdm}}
\end{center}
\end{table}
The remaining couplings are discussed in~\cite{Muhlleitner:2016mzt}.

%%%%%%%%%%%%%%%%%%%%%%%%%%%%%%%%%%%%%%%%%%%%%%%%%%%%%%%
\subsection{The Dark Doublet Phase (DDP)}
In the DDP only one of the two doublets and the singlet acquire VEVs and the $\mathbb{Z}^{(1)}_2$ symmetry forces all the fields in the other
doublet to conserve the darkness parity. The lightest of these dark scalars is a DM candidate.

Assuming that $\Phi_1$ is the SM-like doublet, the vacuum configuration in the DDP is given by
\begin{align}
\left<\Phi_1\right>=\dfrac{1}{\sqrt{2}}\begin{pmatrix}
  0 \\ v
  \end{pmatrix},\qquad
\left<\Phi_2\right>=\begin{pmatrix}
  0 \\ 0
  \end{pmatrix},\qquad
\left<\Phi_S\right>=v_s\,,
\end{align}
where $v\approx$  246 GeV is the electroweak VEV and $v_s \neq 0$ is the singlet VEV. The difference between the non-dark sector
and the SM is that the singlet $\rho_s$ will mix with the CP-even $\rho_1$.
The mass eigenstates $H_i$ ($i=1,2,3$) are obtained from $(\rho_1,\, \rho_2,\, \rho_S)$ via the rotation matrix
\begin{align}
\mathcal{R} = \begin{pmatrix}
\cos\alpha & 0 & \sin\alpha \\  -\sin\alpha & 0& \cos\alpha \\ 0 & 1 & 0
\end{pmatrix}\label{eq:inertdoubletscalarmixingmatrix}.
\end{align}
By convention, we order the visible $H_i$ by ascending mass
\begin{align}
m_{H_1} \le m_{H_2}
\end{align}
and choose the third mass eigenstate $H_D\equiv H_3=\rho_S$.
There is no mixing between the remaining components of the two doublets and therefore
\begin{alignat}{4}
  G^0 &= \eta_1\,,\qquad &A_D &= \eta_2\,,\\
  G^\pm &= \phi_1^\pm\,,\qquad &H^\pm_D &=
  \phi_2^\pm \,.
\end{alignat}
The Goldstone bosons are in the SM-like doublet and the dark charged and dark CP-odd particles are in the inert doublet.\footnote{Note that just like in the IDM  there is no way to tell which of $H_D$ and $A_D$ is the CP-even and which is the CP-odd state. In fact, since both $H_D$ and $A_D$ do not couple to fermions, it is just the $H_D\, A_D \, Z$  coupling that tells us
they have opposite CP. Regardless, we will call $H_D$ CP-even and $A_D$ CP-odd throughout this paper for simplicity.}

In the DDP, the 11 parameters of the N2HDM, \eqref{eq:lagrangepars}, are
expressed through
\begin{equation}
   v\,,\enspace v_s\,,\enspace
  m_{H_{1}}\,,\enspace m_{H_{2}}\,,\enspace m_{H_{D}}\,,\enspace
  m_{A_D}\,,\enspace m_{H^{\pm}_D}\,, \enspace
   \alpha\,,\enspace m_{22}^2\,,\enspace \lambda_2\,,\enspace \lambda_8\,.
  \end{equation}
The explicit parameter transformations are given in Appendix~\ref{app:DDP}.

The couplings of the scalars to the remaining SM particles can be grouped into a \textit{visible sector} consisting of the two neutral CP-even fields $H_1$ and $H_2$
and the \textit{dark sector} with the four scalars $H_D$, $A_D$ and $H^\pm_D$. The coupling modifiers in the visible sector are given by
\begin{equation}
c(H_i(p)) = \frac{\lambda^{(p)}_{i}}{\lambda^{(p)}_{SM}}=\mathcal{R}_{i1}
\end{equation}
where $H_i$ ($i=1,2$) and $p$ stands for a pair of SM particles, provided that there is a corresponding coupling in the SM\@.
$\lambda$ stands for the Feynman rule of the corresponding vertex and
the division by $\lambda_{SM}$ is taken to cancel
identical tensor structures.
Because this visible sector is just the extension of the SM by a real singlet the following sum rules hold:
\begin{align}
\sum\limits_{i=1}^2 c^2(H_i\bar{f}f) = \sum\limits_{i=1}^2 c^2(H_i VV) = 1\,.\label{eq:idp_sumrule}
\end{align}
Finally no FCNC occur at tree-level because only the first doublet couples to fermions.

Due to the unbroken  $\mathbb{Z}^{(1)}_2$ symmetry the dark scalars $H^\pm_D$, $H_D$ and $A_D$ do not couple to either
pairs of fermion or pairs of gauge bosons.
However --- because of the doublet nature of $\Phi_2$
--- there are couplings involving two dark scalars and one vector boson in addition to
 the triple-Higgs couplings $H_i H_D H_D$, $H_i A_D A_D$
and $H_i H^\pm_D H^\mp_D$ that link the dark and the visible sectors.
The trilinear Higgs gauge couplings are dependent on the momenta of the scalars and
there is no SM equivalent with which they could be
normalised.
Adopting the convention in which the momentum $p_{H_D}$ of $H_D$ is incoming, and the momenta $p_{A_D}$ and $p_{H^\pm_D}$ of the scalars $A_D$ or $H_D^\pm$ are outgoing,
we get the following Feynman rules
\begin{alignat}{2}
&\lambda^\mu (H_D, A_D, Z) &&= -\dfrac{\sqrt{g^2 + g'^2}}{2}\, \left(p_{A_D} + p_{H_D}\right)^\mu\,,\\
&\lambda^\mu (H_D, H_D^\pm, W^\mp) &&= \mp \dfrac{i g}{2}\, \left(p_{H_D^\pm} + p_{H_D}\right)^\mu\,.
\end{alignat}
These, and the Feynman rules for the vertices $A_D H_D^\pm W^\mp$, $H_D^\pm
H_D^\mp Z$ and $H_D^\pm H_D^\mp \gamma$ are the same as in the 2HDM
and can be found in Ref.~\cite{Gunion:1989we}.
The triple Higgs couplings are given in appendix~\ref{app:DDP}.\\
%\JWc{I find it weird that we just copy the appendices from~\cite{Engeln:2018mbg}, we should just refer to them there. PF - I AGREE.}

%%%%%%%%%%%%%%%%%%%%%%%%%%%%%%%%%%%%%%%%%%%%%%%%%%%%%%%
\subsection{The Dark Singlet Phase (DSP)}
\label{sec:darksinglet}
In the DSP only the doublets acquire VEVs which means that the $\mathbb{Z}^{(2)}_2$ symmetry is left unbroken.
In turn, only the CP-even fields $\rho_1$ and $\rho_2$ mix and $\rho_S$ is the DM candidate. Now, the
vacuum configuration is
\begin{align}
\left<\Phi_1\right>=\dfrac{1}{\sqrt{2}}\begin{pmatrix}
0 \\ v_1
\end{pmatrix},\qquad
\left<\Phi_2\right>=\dfrac{1}{\sqrt{2}}\begin{pmatrix}
0 \\ v_2
\end{pmatrix},\qquad
\left<\Phi_S\right>=0\,,
\end{align}
where $v_1= v\cos\beta$ and $v_2=v\sin\beta$ and $v$ is the electroweak VEV.
To rotate from the gauge eigenstates $(\rho_1,\, \rho_2,\, \rho_S)$ to
the mass eigenstates we define a rotation matrix compatible with the
usual 2HDM definition,
\begin{align}
\mathcal{R}
=\begin{pmatrix} -\sin\alpha & \cos\alpha & 0 \\
\cos\alpha & \sin\alpha & 0\\
0 & 0 & 1 \end{pmatrix},
\label{eq:DSPmixingmatrix}
\end{align}
where we use the mass ordering
\begin{align}
m_{H_1} \le m_{H_2}\,.
\end{align}
$H_3=\rho_S$ is the dark scalar $H_D$.
The CP-odd and charged eigenstates are obtained exactly like in the 2HDM case, that is,
\begin{alignat}{6}
G^0 &= \eta_1\cos\beta&&+\eta_2\sin\beta\,,\qquad &A &=-\eta_1\sin\beta&&+\eta_2\cos\beta\,,\\
G^\pm &= \phi_1^\pm\cos\beta&&+\phi_2^\pm\sin\beta\,,\qquad &H^\pm &=-\phi_1^\pm\sin\beta&&+\phi_2^\pm\cos\beta\,.
\end{alignat}

In the DSP, the 11 parameters of the N2HDM, \eqref{eq:lagrangepars}, are expressed in terms of the input parameters
as
\begin{alignat}{2}
v\,,\enspace \tan\beta\,,\enspace
m_{H_{1}}\,,\enspace m_{H_{2}}\,,\enspace m_{H_{D}}\,,\enspace
m_{A}\,,\enspace m_{H^{\pm}}\,, \alpha\,,\enspace \lambda_6\,,\enspace \lambda_7\,,
\enspace \lambda_8\,,\notag
\end{alignat}
and the explicit transformation of the parameters can be found in Appendix~\ref{app:DSP}.

Regarding the Higgs couplings, the singlet field $\rho_S$ does not couple to SM particles nor does it mix with the remaining CP-even scalar fields $\rho_1$ and $\rho_2$.
Hence the $H_1$ and $H_2$ couplings to the SM particles are just the
2HDM Type I ones and can be found in Table~\ref{tab:dsp_fermioncouplings}.
\begin{table}[t]
\centering
\begin{tabular}{lcr}
\toprule
		& $c(H_i \bar{f}f)$			& \multicolumn{1}{c}{$c(H_i VV)$} \\
\midrule
$H_1$ 	& $\cos\alpha/\sin\beta$	& $-\sin\left(\alpha-\beta\right)$		\\
$H_2$ 	& $\sin\alpha/\sin\beta$	& $ \cos\left(\alpha-\beta\right)$		\\
\bottomrule
\end{tabular}
\caption{Yukawa and gauge boson coupling modifiers for the CP-even Higgs bosons $H_i$ $(i =1,2)$ in the DSP.}
\label{tab:dsp_fermioncouplings}
\end{table}
The only additional couplings are the triple-Higgs couplings $H_i H_D H_D$ $(i =1,2)$, which allow for the decay of the light and heavy CP-even Higgs boson into
DM if kinematically possible. These interactions have the form
\begin{alignat}{2}
g (H_i H_D H_D) = \,\dfrac{\partial \mathcal{L}}{\partial H_i\partial H_D\partial H_D} =
   \lambda_7 v \cos\beta \mathcal{R}_{i1}
 + \lambda_8 v \sin\beta \mathcal{R}_{i2}\,,
\end{alignat}
where $\mathcal{R}_{ij}$ is the $ij$ element of the mixing matrix in
\eqref{eq:DSPmixingmatrix}.

%%%%%%%%%%%%%%%%%%%%%%%%%%%%%%%%%%%%%%%%%%%%%%%%%%%%%%%
\subsection{The Fully Dark Phase (FDP)}
\label{sec:SMP}
In the FDP only one
doublet acquires a VEV\@. This means that both $\mathbb{Z}^{(1)}_2$ and
$\mathbb{Z}^{(2)}_2$ remain unbroken and we have two DM candidates corresponding
to the two different dark parities. Because all other neutral fields belong to
one of the dark phases, the SM-like Higgs is just the one from the doublet with
a VEV. There is no mixing in the scalar sector, such that
$\mathcal{R}=\mathbb{1}_{3\times3}$ in the basis
\begin{equation}
\begin{pmatrix}H_\text{SM}\\H^D_D\\H^S_D\end{pmatrix} = R \begin{pmatrix}\rho_1\\\rho_2\\\rho_S\end{pmatrix}
\end{equation}
where we denote by $H_{D}^D$ ($H_{D}^S$) the CP-even, dark scalar from the
doublet (singlet).
Hence, $H_\text{SM}$ has exactly the same couplings to SM particles as in the SM. The only difference
relative to the SM are the couplings between the Higgs and the dark matter candidates stemming
from the Higgs potential. There is, however, a difference in the SM
Higgs radiative decays and in particular $H_\text{SM} \rightarrow \gamma\gamma$ where the contribution from the dark charged Higgs loops can
significantly change $\Gamma (H \rightarrow \gamma\gamma)$.
In the FDP, the 11 parameters of the N2HDM, \eqref{eq:lagrangepars}, are expressed through
\begin{alignat}{2}
v\,,\enspace  m_{H_{SM}}\,,\enspace
m_{H_{D}^D}\,,\enspace m_{H_{D}^S}\,,\enspace m_{A_{D}}\,,\enspace
m_{H_D^{\pm}}\,,\enspace m_{22}^2\,,\enspace m_S^2\,,\enspace \lambda_2
\,,\enspace \lambda_6 \,,\enspace \lambda_8\,.
\end{alignat}

%%%%%%%%%%%%%%%%%%%%%%%%%%%%%%%%%%%%%%%%%%%%%%%%%%%%%%%

\section{Neutral Vacua Stability}
\label{sec:bil}

The existence of several possible vacua, wherein different discrete symmetries of the model
are broken by the vevs, raises the possibility of coexisting minima. Namely, is it guaranteed
that once we find a given minimum -- corresponding to one of the phases defined in
section~\ref{sec:model} -- that this minimum is the {\em global} one? Or may
deeper neutral minima exist, raising the possibility of tunnelling between minima?
In order to answer this question one must compute the values of the potential at different
coexisting vacua and compare them.
In the context of charge breaking vacua in the N2HDM the authors of the present work analysed
this possibility in Ref.~\cite{Ferreira:2019iqb}
(see also \cite{Basler:2018cwe, Basler:2019iuu}). We now undertake a similar study for
coexisting neutral vacua following earlier numerical studies in Refs.~\cite{Muhlleitner:2016mzt, Engeln:2018ywp}.

To begin with, some generic considerations:
\begin{itemize}
\item In all that follows, we will always assume that two stationary points, corresponding
to different phases of the model, coexist. This means that, for some set of parameters of
the potential, we are assuming that the minimization conditions of the potential admit
two solutions, with different values for the vevs.
\item Since we will be comparing the values of the potential at different phases of the model,
we must distinguish between the vevs $v_1$, $v_2$ and $v_s$ defined previously. Therefore,
each vev will, for the purposes of this section alone, be tagged with a superscript to
specify which neutral phase is being discussed. The vevs of the Broken Phase (BP), for instance,
will be tagged with a ``B" -- $v_1^B$, $v_2^B$ and $v_s^B$ -- whereas those of the Dark Doublet
Phase (DDP) will carry a ``D" -- $v_1^D$ and $v_s^D$. The complete correspondence can be found
in Table~\ref{tab:vevs}. Likewise, scalar masses at different phases will carry the same subscript
%
%\item In the phenomenological study our minima are either global or if local, the tunnelling time to a deeper
%minimum is larger than the age of the Universe.
%such that    We will not study tunneling times between different vacua in this work. Similar studies have
%recently been undertaken for the 2HDM~\cite{Branchina:2018qlf}, the MSSM~\cite{Hollik:2018wrr} and
%the N2HDM~\cite{Ferreira:2019iqb}, but for the current work we always made sure (numerically) that
%our minima are the global extrema of the potential \textcolor{red}{(JONAS IS THIS CORRECT?)}
%
\end{itemize}

\begin{table*}[t]
  \begin{center}
  \renewcommand*{\arraystretch}{1.5}
  \begin{tabular}{lc}
  \toprule
   Phase  & vevs \\
   \midrule
    BP & $v_1^B$, $v_2^B$, $v_s^B$ \\
    DDP & $v_1^D$, $v_s^D$ \\
    DSP & $v_1^S$, $v_2^S$ \\
    FDP & $v_1^F$ \\
  \bottomrule
  \end{tabular}
  \end{center}
  \caption{Naming convention for the vevs at stationary points of different phases. Only non-zero vevs are
    shown.}
  \label{tab:vevs}
  \end{table*}

In order to compare the values of the potential at different vacua we will deploy a bilinear
formalism similar to the one employed for the
2HDM~\cite{Velhinho:1994np,Ferreira:2004yd,Barroso:2005sm,Nishi:2006tg,Maniatis:2006fs,Ivanov:2006yq,
  Barroso:2007rr,Nishi:2007nh,Maniatis:2007vn,Ivanov:2007de,Maniatis:2007de,Nishi:2007dv,
  Maniatis:2009vp,Ferreira:2010hy}.
In this approach, bilinears are several gauge-invariant quantities, quadratic in the fields,
and the potential, expressed in terms of these variables, becomes a quadratic polynomial.
The minimisation of the potential is greatly simplified, and geometrical properties of these
bilinears permit a detailed analysis of symmetries of the potential and its vacuum structure.
This formalism has been
adapted to study the vacuum structure of other models, such as the three-Higgs doublet
model~\cite{Ivanov:2010ww,Ivanov:2014doa,Ivanov:2018jmz}, the doublet-singlet model~\cite{Ferreira:2016tcu},
the N2HDM~\cite{Ferreira:2019iqb} and the Higgs-triplet
model~\cite{Ferreira:2019hfk}. We now give a brief overview
of the technique: let us define vectors $A$ and $X$ and
a matrix $B$ as
\be
X = \frac{1}{2}\,\left(\begin{array}{c} v_1^2 \\ v_2^2 \\ v_1 v_2 \\ v_s^2
  \end{array}\right)\,,\quad
A = \left(\begin{array}{c} m_{11}^2 \\ m_{22}^2 \\ 0  \\ m^2_S
  \end{array}\right)\,,\quad
B = \left(\begin{array}{cccc} \lambda_1 & \lambda_3 & 0      & \lambda_7 \\
    \lambda_3         & \lambda_2 & 0                        & \lambda_8 \\
    0                 & 0         & 2(\lambda_4 + \lambda_5) & 0         \\
    \lambda_7         & \lambda_8 & 0                        & \lambda_6\end{array}\right) \, .
\label{eq:def}
\ee
The value of the potential of \eqref{eq:scalpot} at any of the phases (corresponding to a stationary point (SP))
we consider in this work can be then be expressed as
\be
V_{SP}  =  A^T\,X_{SP} \,+\, \frac{1}{2}\,X_{SP}^T \,B\,X_{SP}\,,
\label{eq:vpr}
\ee
with the vector $X$ evaluated at the stationary point,
and it can easily be shown that, due to the minimisation conditions, one has
\be
V_{SP}  = \frac{1}{2}\,A^T X_{SP}  = -\,\frac{1}{2}\,X_{SP}^T B X_{SP}\,.
\label{eq:Vmin}
\ee
The bilinear formalism also requires that we define the following vector
\be
V^\prime_{SP} = \frac{\partial V}{\partial X^T}  =
A\,+\,B\,X_{SP}\, .
\label{eq:defVl}
\ee
In order to  illustrate the technique we will now show how to apply the formalism to one of the cases we are
interested in, detailing the several steps needed to reach a formula comparing the depth of the potential
 at two different phases. We will then simply present the results obtained for all the other cases
 without demonstration.\footnote{We leave it as an exercise to the reader, contributing this way to the fight against the state of boredom that hit
 particle physicists all around the globe.}

 Suppose the N2HDM potential of~\eqref{eq:scalpot} has two stationary points, corresponding to
 the phases BP and DDP, defined in section~\ref{sec:model}. Then, the vectors $X$ and $V^\prime$
 have the following expressions for each phase: for the Broken Phase,
\be
X_{BP}  = \frac{1}{2}\,
\left(\begin{array}{c} (v_1^B)^2 \\ (v_2^B)^2 \\ v_1^B v_2^B \\ (v_s^B)^2 \end{array}\right)\
\,,\quad
V^\prime_{BP}  =  A\,+\,B\,X_{BP} =\,-\,\frac{\lambda_4 + \lambda_5}{2}\,
\left(\begin{array}{c} (v_2^B)^2 \\ (v_1^B)^2 \\ -2 v_1^B v_2^B \\ 0\end{array}\right)\, ,
\label{eq:xvBP}
\ee
and for the Dark Doublet Phase,
\be
X_{DDP}  = \frac{1}{2}\,
\left(\begin{array}{c} (v_1^D)^2 \\ 0 \\ 0 \\ (v_s^D)^2 \end{array}\right)\
\,,\quad
V^\prime_{DDP}  =  A\,+\,B\,X_{DDP} =\,-\,(m^2_{H^\pm})^{D} \,
\left(\begin{array}{c} 0 \\ 1 \\ 0 \\ 0 \end{array}\right)\, ,
\ee
where the charged scalar mass at the DDP extremum is given by
\be
(m^2_{H^\pm})^{D}\,=\,m^2_{22} \,+\,\frac{1}{2} \lambda_3\,(v_1^D)^2
\,+\,\frac{1}{2} \lambda_8\,(v_s^D)^2\,.
\ee
We then compute the following product between vectors:
\begin{align}
  X^T_{BP} V^\prime_{DDP} & = X_{BP}^T\,A\,+\, X^T_{BP} B X_{DDP} =
  2 V_{BP}\,+\, X^T_{BP} B X_{DDP} \nonumber \\
  & = -\frac{\lambda_4 + \lambda_5}{4} \,(v_2^B)^2 \,(v_1^D)^2
\end{align}
where in the second line we used the result from  \eqref{eq:Vmin}. Likewise, we obtain
\begin{align}
  X^T_{DDP} V^\prime_{BP} & = X_{DDP}^T\,A\,+\, X^T_{DDP} B X_{BP} =
  2 V_{DDP}\,+\, X^T_{DDP} B X_{BP} \nonumber \\
  & = \frac{1}{2} \,(v_2^B)^2 \,(m^2_{H^\pm})^{D}\,.
\end{align}
Since the matrix $B$ is symmetric we will have $X^T_{BP} B X_{DDP} = X^T_{DDP} B X_{BP}$, and therefore,
subtracting the two equations above one from another
we obtain, after some intermediate steps that we skip for brevity,
\be
V_{BP}\,-\,V_{DDP}\,=\,\frac{1}{4} \,(v_2^B)^2 \,(m^2_{H_D})^{D}\,,
\label{eq:difV1}
\ee
where $(m^2_{H_D})^{D}$ is the squared scalar mass corresponding of the
real, neutral component of the doublet $\Phi_2$ in the DDP phase (see
Appendix~\ref{app:DDP}). What \eqref{eq:difV1} shows us is that, if the Dark
Doublet Phase is a minimum, then all of the squared scalar masses therein
computed will perforce be positive and then one will necessarily have
\be
V_{BP}\,-\,V_{DDP}\,>\,0\qquad \text{if DDP is a minimum.}
\ee
Therefore, if DDP is a minimum, any stationary point corresponding to the Broken Phase will
necessarily lie {\em above} that minimum.

Following similar steps we can obtain the relations between the BP potential value and the remaining
phases, namely
\begin{align}
V_{BP}\,-\,V_{DSP} & = \, \frac{1}{4} \,(v_s^B)^2 \,(m^2_{H_D})^{S}\,, \\
 & \nonumber \\
 V_{BP}\,-\,V_{FDP} & = \, \frac{1}{4} \,(v_2^B)^2 \,(m^2_{H_D^D})^{F}\,+\,
 \frac{1}{4} \,(v_s^B)^2 \,(m^2_{H_D^S})^{F}\,,
\label{eq:vfdp}
\end{align}
where the $m^2$ are physical scalar masses at the given phases. From these
equations one draws analogous conclusions to the case with coexisting BP and DDP
phases.

The above does not answer the question of whether a local BP minimum could coexist with a
deeper DDP, DSP or FDP minimum, however.
%\textcolor{orange}{You mean it could coexist as a metastabel minimum?}
We will now show that
%But previous similar expressions and results for the
%2HDM and other models (Higgs-triplet model, complex singlet model, N2HDM) lead us to
%argue that this almost certainly \textcolor{orange}{What do you mean
 % to say by 'almost certainly'?} means that when  a DDP, DSP or FDP minimum exists, a
any
BP stationary point will necessarily be a saddle point:
%. And in fact this is what we find, as
%we will now show:
in the BP phase, the real neutral components of both doublets, $\rho_1$ and $\rho_2$,
mix with the singlet component field $\rho_S$,
leading to a $3\times 3$ scalar mass matrix for the CP-even
scalars, $M_{\text{scalar}}^2$ (see
section~\ref{sec:model}).
%\textcolor{orange}{We do not really introduce this mass matrix there.}
  It is possible to show that there
is an alternative way of writing~\eqref{eq:difV1}, to wit
\be
V_{BP}\,-\,V_{DDP}\,=\,\frac{1}{4} \,(v_2^B)^2 \,(m^2_{H_D})^{D}\,=\,
-\,\frac{1}{8(\lambda_1 \lambda_6 - \lambda_7^2)}\,\frac{(v_2^B)^2}{(v_1^B)^2 (v_s^B)^2}
\,\mbox{det}\left(M_{\text{scalar}}^2\right)_B\,,
\ee
where we added the subscript ``B" to the determinant to emphasise
that these scalar masses
are evaluated at the BP extremum. It can be shown that, if the DDP phase is a minimum,
then one must have $\lambda_1 \lambda_6 - \lambda_7^2 > 0$ (to do this one must look at the
DDP scalar mass matrix, see Appendix~\ref{app:DDP}).
 Therefore, if the DDP is a minimum then
$V_{BP}\,-\,V_{DDP}\,>\,0$ and $\mbox{det}\left(M_{\text{scalar}}^2\right)_B < 0$ ---
which means that at least one BP squared scalar mass is negative. Since $\left(M_{\text{scalar}}^2\right)_B$
 is a matrix with positive diagonal entries some of its minors
 are guaranteed to be positive ---
 and therefore we conclude that at least one of its eigenvalues is positive. Therefore,
 if the DDP is a minimum, the broken phase BP is {\em a saddle
   point}.  Reversely, if the BP {\em is} a minimum, then one will have
$V_{BP}\,-\,V_{DDP}\,<\,0$  {\em and}
the DDP extremum cannot be a minimum, and indeed it
 can be shown to be a saddle point.
Analogous expressions can be found for the comparison between the BP and the other neutral phases.
Thus one may conclude the following:
%\textcolor{orange}{You repeat
 % here insights on BP from above but not the insights on DSP, DSP and
 % FDP from above. This is confusing.}
%
\begin{itemize}
\item If any of the phases DDP, DSP and FDP is a minimum, then any stationary point
of the BP lies necessarily above that minimum, and is a {\em saddle point}.
\item If there is a minimum of the  scalar potential in the BP, then any
stationary points of the DDP, DSP and FDP are necessarily saddle points
and lie {\em above} the BP minimum.
\end{itemize}

We can easily find the relationship between the depths of the potential at DDP and DSP
phases -- analogous calculations lead us to
\be
V_{DSP}\,-\,V_{DDP}\,=\,\frac{1}{4} \,(v_2^S)^2 \,(m^2_{H_D})^{D}\,-\,
\frac{1}{4} \,(v_s^D)^2 \,(m^2_{H_D})^{S}\,,
\ee
where we see that now, even if either the DSP or the DDP, or both, are minima, there is
no assurance whatsoever that it is the deepest minimum. In fact, the above expression, from
previous 2HDM and N2HDM experience, implies that DSP and DDP minima
can coexist and either can be the deepest minimum, depending on the choice of parameters of the potential.

Finally, one can analyse the FDP phase. We already saw (\eqref{eq:vfdp} above) that an FDP
minimum implies that any BP extrema lies above it. When we compare FDP stationary points
with DDP and DSP ones, we obtain the following expressions,
\begin{align}
V_{DDP}\,-\,V_{FDP} & = \, \frac{1}{4} \,(v_s^D)^2 \,(m^2_{H_D^S})^{F}\,, \nonumber \\
 & \nonumber \\
 V_{DSP}\,-\,V_{FDP} & = \, \frac{1}{4} \,(v_2^S)^2 \,(m^2_{H_D^D})^{F}\,,
\end{align}
which again show that, if the FDP is a minimum, any extrema
corresponding to the phases
DDP and DSP necessarily will lie {\em above it} --- and as happened for the BP phase, it
 can be shown
that in that case the DDP and DSP phases would not be minima, but rather saddle points.
Likewise, the existence of DDP/DSP minima would imply
that any FDP extremum would lie above it, and it
would be a saddle point.
From~\eqref{eq:vfdp} and these results we can therefore safely conclude that {\em a minimum
in the FDP is deeper than any other extrema for different neutral phases}.

To summarise, then:
\begin{itemize}
\item If a BP minimum exists it is the global minimum of the theory. All other
stationary points corresponding to different phases lie above it and
are saddle points.
\item Likewise for the existence of a FDP minimum --- if it exists it is global and all other
other
stationary points corresponding to different phases lie above it and
 are saddle points.
\item However, minima of the DDP and DSP can coexist in the potential, and neither is guaranteed
to be deeper than the other. If there is a minimum DDP or DSP, any BP or FDP extrema are
saddle points above it.
\end{itemize}
This last point recalls the coexistence of minima which break the same symmetries in the
2HDM~\cite{Ivanov:2007de}. Although in the DDP and DSP phases different symmetries are broken,
the symmetry of these models after spontaneous symmetry breaking is very similar in both models,
as a $\mathbb{Z}_2$ symmetry is left unbroken by the vacuum in both models.

We therefore were able to find general statements about the N2HDM vacuum structure in an
analytical manner. Stability of BP and FDP phases is assured, but numerical checks need to be performed
on DDP and DSP ones in order to verify whether a minimum of these phases is the global one.
A final note about having set $m_{12}^2=0$. As already discussed in~\cite{Ferreira:2019iqb}, if $m_{12}^2 \neq 0$
the result that compares the BP with the DSP no longer
holds.
%\textcolor{red}{Can you remind me why not?}
Let us now proceed to the numerical analysis of the several phases.

%%%%%%%%%%%%%%%%%%%%%%%%%%%%%%%%%%%%%%%%%%%%%%%%%%%%%%%
\section{Parameter Scans and Constraints \label{sec:scans}}

All phases of the N2HDM have been implemented in the {\tt ScannerS}
code~\cite{Coimbra:2013qq,SCANNERS2} to perform parameter scans and in the
{\tt N2HDECAY} code~\cite{Muhlleitner:2016mzt,Engeln:2018mbg} to calculate all Higgs branching
ratios and decay widths including state-of-the-art higher-order QCD corrections
and off-shell decays. Electroweak corrections, which --- in contrast to the QCD
corrections --- cannot be taken over from the SM, have been consistently
neglected.\footnote{While there exists a the code {\tt
    ewN2HDECAY} \cite{Krause:2019oar} that calculates the electroweak corrections to
  the on-shell and not loop-induced decays of the neutral N2HDM Higgs
  bosons in the broken phase it has not been adapted yet to the dark phases
discussed in this paper.} Since we only consider type I Yukawa sectors --- where the
effective couplings of each visible Higgs boson to all fermions are equal ---
the scalar production cross sections are easily obtained for all phases from the
corresponding SM ones --- calculated using {\tt SusHi
v1.6.1}~\cite{Harlander:2012pb,Harlander:2016hcx} (see also~\cite{Harlander:2013qxa}).

The parameter points generated using \texttt{ScannerS} in each model are in
agreement with the most relevant theoretical and experimental constraints.
Theoretical constraints include that the potential is bounded from below and
that perturbative unitarity holds~\cite{Muhlleitner:2016mzt}. We further require stability of the EW vaccuum, and also allow for metastability using the numerical procedure described in Refs.~\cite{Hollik:2018wrr,Ferreira:2019iqb},
provided the tunnelling time to a deeper minimum is larger than the age of the Universe.
The SM-like Higgs boson mass is taken to be~\cite{Aad:2015zhl}
\beq
m_{h_{125}} = 125.09 \; \mbox{GeV}\;,
\eeq
and to preclude interference with other Higgs signals we force any non-dark neutral
scalar to be outside the $m_{h_{125}} \pm 5$~GeV mass window. Any of
the visible CP-even Higgs bosons
can be the discovered one.

Compatibility with electroweak precision data is imposed by a 95\% C.L.~exclusion
limit from the electroweak precision observables $S$, $T$ and $U$~\cite{Peskin:1991sw} using
the formulae in Refs.~\cite{Grimus:2007if,Grimus:2008nb} and the fit result of
Ref.~\cite{Haller:2018nnx}. In the BP and DSP
we also consider constraints from charged-Higgs mediated contributions to $b$-physics observables~\cite{Haller:2018nnx}.

Constraints from Higgs searches are taken into account using the combined
95\% C.L.~exclusion bound constructed by
\texttt{HiggsBounds}-5.7.1~\cite{Bechtle:2008jh,Bechtle:2011sb,Bechtle:2013wla}
including LEP, Tevatron and LHC results. The measurements of the $h_{125}$
properties at the LHC are included through the use of
\texttt{HiggsSignals}-2.4.0~\cite{Bechtle:2013xfa}, where a $\Delta\chi^2<6.18$
cut relative to the SM is used.

In the dark phases, additional constraints from DM observables are considered. The relic density and
direct detection cross sections are calculated using
\texttt{MicrOMEGAs}-5.0.9~\cite{Belanger:2006is,Belanger:2008sj,
Belanger:2010gh,Belanger:2013oya,Belanger:2014vza,Barducci:2016pcb,Belanger:2018mqt}.
This calculation correctly accounts for the two-component DM in the FDP. The model-predicted relic density is
required not to oversaturate the observed relic abundance~\cite{Aghanim:2018eyx} by more than $2\sigma$.
Additionally, the direct detection bound by the \textsf{Xenon1t} experiment~\cite{Aprile:2018dbl} is imposed.

%
%%%%%%%%%%%%%%%%%%%%%%%%%%%%%%%%%%%%%%%%%%%%%%%%%%%%%%%
\begin{figure}[t]
  \centering
\includegraphics[width=0.69\linewidth]{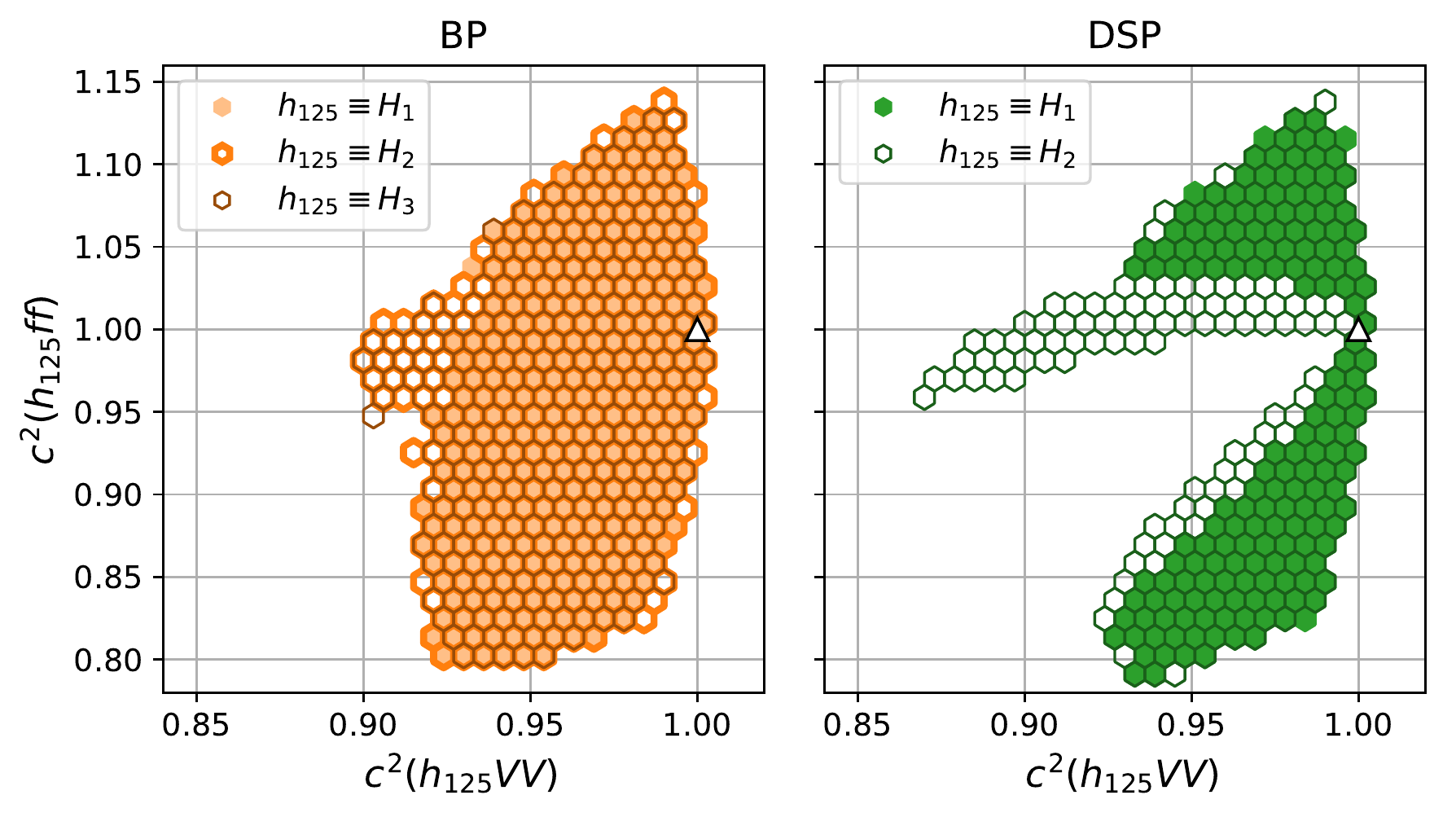}
  \caption{Coupling modifiers relative to the SM for the couplings of $h_{125}$ to fermions, $c(h_{125} \bar f f)$, and to gauge bosons, $c(h_{125} VV)$,
  for the Broken Phase (left) and for the Dark Singlet Phase (right). The white triangle indicates the SM value.
    }\label{fig:coup}
\end{figure}
%%%%%%%%%%%%%%%%%%%%%%%%%%%%%%%%%%%%%%%%%%%%%%%%%%%%%%%
%

Let us now understand what are the present bounds on the Higgs couplings modifiers.
In Fig.~\ref{fig:coup} we present the squared coupling modifiers to fermions and to gauge bosons of the $125\;\text{GeV}$ Higgs boson $h_{125}$. We show the Broken
Phase (left) and the Dark Singlet Phase (right).
Due to unitarity, the effective coupling to gauge bosons cannot exceed 1. We also show the differences in the allowed parameter space when
considering the different CP-even scalars as the $h_{125}$.
In both phases, we see that lower values of $c^2(h_{125}VV)$ are allowed if $h_{125}$ is not the lightest of the $H_i$. This is the result of more freedom in $\mu_{\gamma\gamma}$ for light spectra --- in particular for light charged Higgs masses. We will explain the origin of this behaviour below, when we discuss $\mu_{\gamma\gamma}$ as a distinguishing factor between the phases.
We do not show the corresponding plots for the other two phases since they are trivial. In the DDP the two effective couplings are always equal and constrained to the experimentally allowed range
\begin{equation}
  0.87 < c^2(h_{125}ff)=c^2(h_{125}VV)<1\,,
\end{equation}
while in the FDP, both couplings take exactly their SM values.

%%%%%%%%%%%%%%%%%%%%%%%%%%%%%%%%%%%%%%%%%%%%%%%%%%%%%%%
\begin{figure}[t]
  \centering
\includegraphics[width=0.45\linewidth]{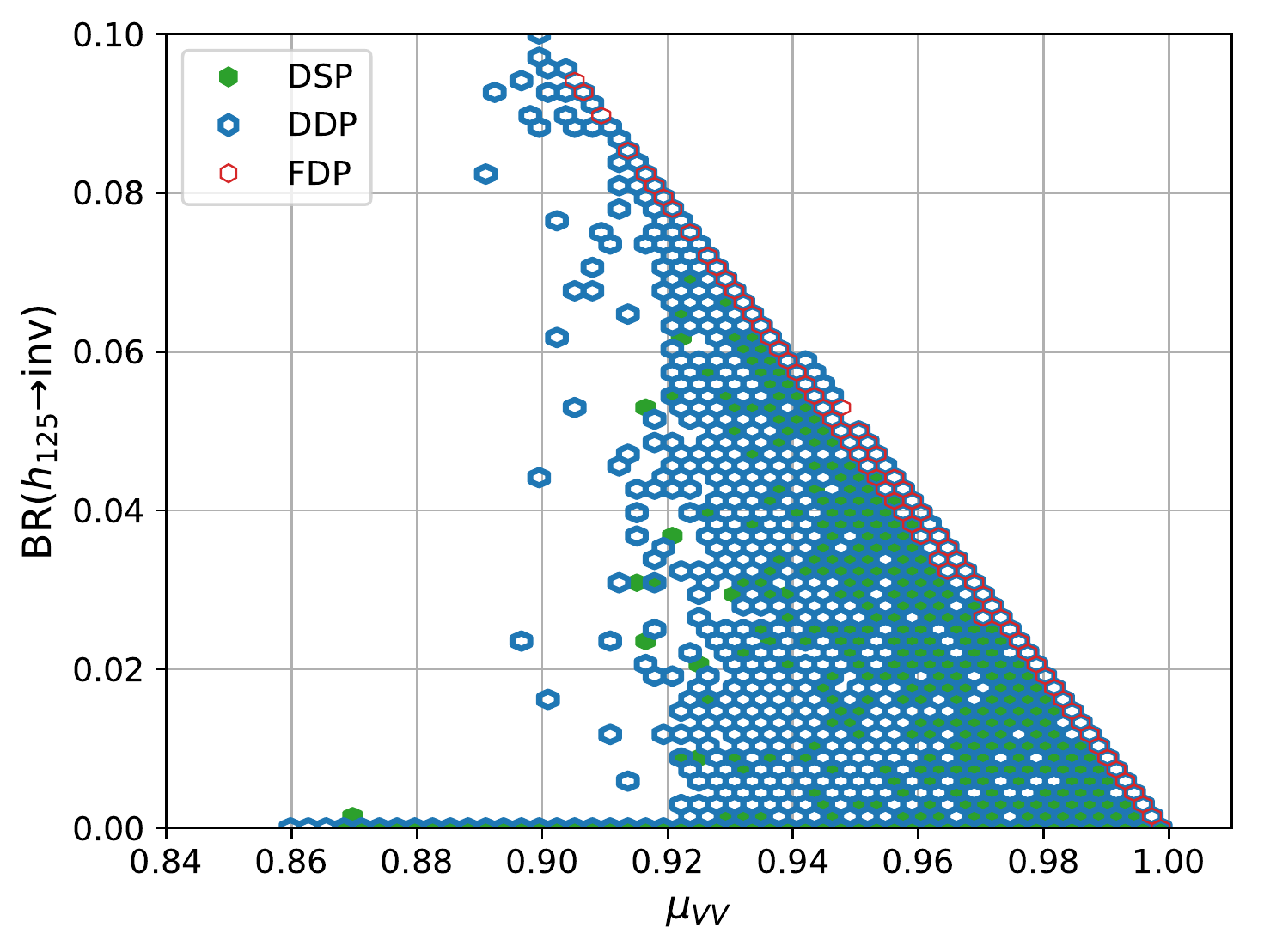}
  \caption{Branching ratio of $h_{125}$ to DM particles vs. $\mu_{VV}$ for the three DM phases.}\label{fig:brinv}
\end{figure}
%%%%%%%%%%%%%%%%%%%%%%%%%%%%%%%%%%%%%%%%%%%%%%%%%%%%%%%
%
In Fig.~\ref{fig:brinv} we show the branching ratio of $h_{125}$ to DM particles vs.  the quantity
\begin{equation}
  \mu_{VV}=\frac{\sigma(pp\to h_{125}\to ZZ)}{\sigma_\text{SM}(pp\to h_{125}\to ZZ)} = \frac{\sigma(pp\to h_{125}\to W^+W^-)}{\sigma_\text{SM}(pp\to h_{125}\to W^+W^-)},
\end{equation}
for the three dark phases.
The maximum allowed value of the branching ratio of the Higgs decaying to DM particles is below
10\% in all phases. The present experimental bound on BR$(h_{125} \to \text{invisible})$ is about
26\%~\cite{Aaboud:2019rtt}. This means that indirect constraints on BR$(h_{125} \to \text{invisible})$
from the Higgs rate measurements are significantly stronger than those from direct searches for invisible decays of $h_{125}$.

Let us now move to the DM constraints. The analysis of the DM phases are the main goal
of this study.  Therefore, we need to make sure
that the DM candidates are compatible with the corresponding
experimental constraints.
The Planck space telescope~\cite{Aghanim:2018eyx} maps the
anisotropies in the cosmic microwave background
radiation.
%\textcolor{red}{What do you mean to say in the previous sentence?}
We force our points to have a relic density of cold dark matter within or below the
$2\times 1\sigma$ band of the experimental fit value
\begin{align}
(\Omega_c h^2)_{\text{exp}} =  0.1200 \pm 0.0012\,.\label{eq:relic-density_Planck}
\end{align}
Hence, points with an over-abundance of DM are excluded. These models are also constrained by
DM direct detection. The most recent results
are the ones from the  \texttt{XENON1T}
experiment~\cite{Aprile:2017iyp} a dual phase (liquid-gas) Xenon time
projection chamber. Because no signal
has been observed so far, constraints in the plane DM-nucleon cross
section vs. DM mass are obtained.
Since the \texttt{XENON1T} bound is obtained assuming a relic density equal to
\eqref{eq:relic-density_Planck} and we allow
for smaller values of the relic densities, the impact of the DM abundance on direct detection
 measurements is taken into account by considering a normalised scattering cross section
  $\hat{\sigma}_{DM-N}$, given by
\begin{align}
\hat{\sigma}_{\text{DM-N}} =&\enspace \sigma_{\text{DM-N}}  \frac{\Omega_c h^2}{(\Omega_c h^2)_{\text{exp}}}\,,\label{eq:normed_scattering_cr}
\end{align}
where $\sigma_{\text{DM-N}}$ and $\Omega_c h^2$ are the values calculated for a given parameter set.
%Consequently, we demand that \textcolor{red}{What is
%  $\sigma_{\text{max}}$? And what is $M_{\text{LDP}}$?}
%\begin{align}
%\hat{\sigma}_{\text{DM-N}} \leq \sigma_{\text{max}}(M_{\text{LDP}})\,.
%\end{align}

%
%%%%%%%%%%%%%%%%%%%%%%%%%%%%%%%%%%%%%%%%%%%%%%%%%%%%%%%
\begin{figure}[tp]
  \centering
\includegraphics[width=0.69\linewidth]{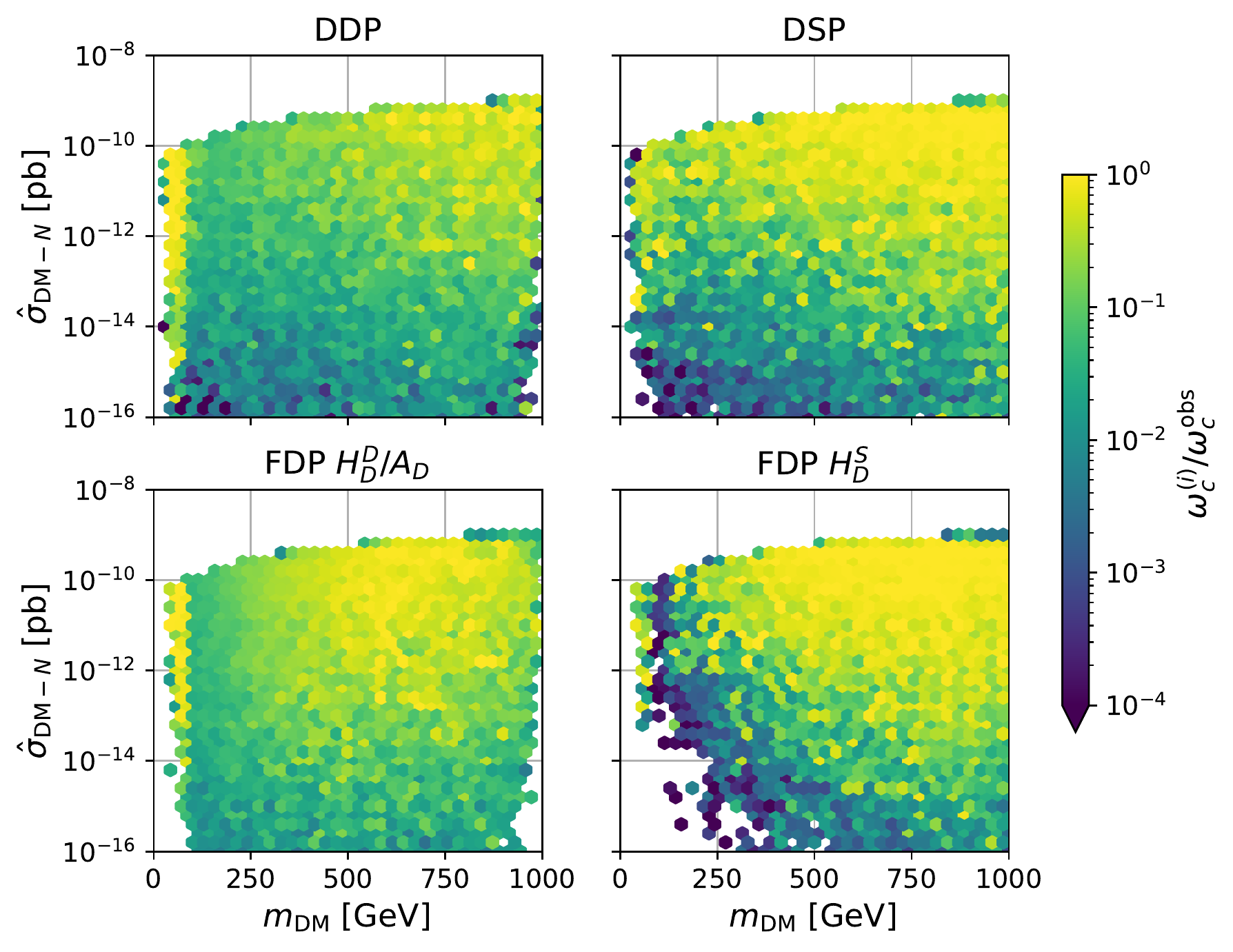}
  \caption{Nucleon-DM cross section, $\hat{\sigma}_{\text{DM-N}}$, as a function of the DM mass with all the constraints previously discussed. The colour code
  represents the fraction of the DM relic density, where the upper
  limit is the central value measured by Planck plus $2\times
  1\sigma$. On the left upper plot we show the DDP and take the
  lightest dark sector particle to be the dark matter candidate; on the
  upper right we show the DSP with the corresponding dark matter candidate. The lower plots show the
  FDP:  here the DM particle is either $H_D^D$ or $A_D$ on the left and $H_D^S$ on the right; note that since the two symmetries
  are conserved separately there are always two dark matter candidates in the FDP.
    }\label{fig:ma}
\end{figure}
%%%%%%%%%%%%%%%%%%%%%%%%%%%%%%%%%%%%%%%%%%%%%%%%%%%%%%%
%

\begin{figure}[tp]
  \centering
\includegraphics[width=0.6\linewidth]{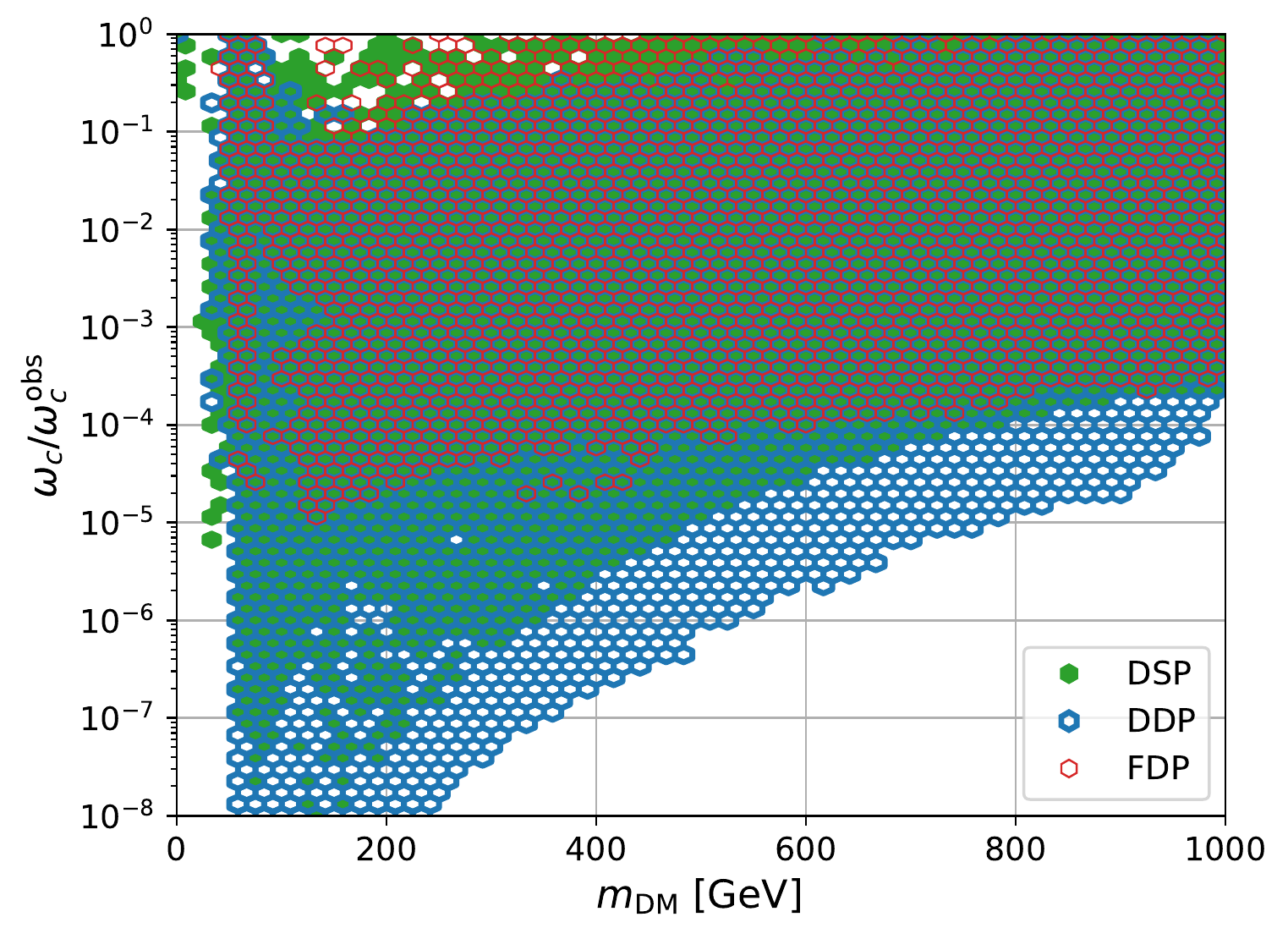}
  \caption{Fraction of the DM relic density as a function of the DM mass for the three DM phases. For the FDP,
  we show the mass of the dark matter candidate that gives the largest
  contribution to the relic density.
      }\label{fig:ome}
\end{figure}

In Fig.~\ref{fig:ma} we  present the Nucleon-DM cross section,
$\hat{\sigma}_{\text{DM-N}}$, as a function of the DM mass with all the
constraints previously discussed. The colour code represents the fraction of the
DM relic density where the upper limit is the central value measured by Planck
plus $2\times 1\sigma$. Regarding direct detection it is clear that plenty of
parameter points  will survive all the way down to the neutrino
floor~\cite{Billard:2013qya} --- which for the mass range in question  is of the
order of $10^{-12}$ pb.  As for saturating the relic density --- allowing
therefore that DM is fully explained within the model --- we now refer to
Fig.~\ref{fig:ome} for clarity. In the figure we see that except for the DDP,
the other phases have points for which $\Omega_c h^2 = (\Omega_c h^2)_{\text{exp}}$
for all values above 125/2~GeV. The DDP has
a DM mass region between about 100 and 500 GeV where did not find any
parameter points that saturate the relic density and extra DM candidates are
needed. This is in line with previous results (see refs.~\cite{Arhrib:2013ela,
Ilnicka:2015jba, Belyaev:2016lok, Kalinowski:2018ylg}) where it was reported
that for the Inert doublet Model, the dark matter relic density cannot be
saturated for DM masses between about 75 and 500 GeV.

%
%%%%%%%%%%%%%%%%%%%%%%%%%%%%%%%%%%%%%%%%%%%%%%%%%%%%%%%
%%%%%%%%%%%%%%%%%%%%%%%%%%%%%%%%%%%%%%%%%%%%%%%%%%%%%%%

\section{The different phases at the LHC and future colliders}
\label{sec:LHC}

The different phases of the N2HDM lead to different phenomenology at
the LHC and at future colliders. There are obvious differences that
would immediately exclude some of them. The discovery
of a charged Higgs boson would immediately exclude the DDP and the
FDP\@. The discovery of three extra neutral scalars in the visible sector would exclude all
phases except the broken phase.
However, the best chances we have to probe the different phases are the 125 GeV
Higgs rates measurements and perhaps the search for an extra neutral scalar.

\subsection{$h_{125}$ coupling measurements}

%
%%%%%%%%%%%%%%%%%%%%%%%%%%%%%%%%%%%%%%%%%%%%%%%%%%%%%%%
\begin{figure}[t]
  \centering
\includegraphics[width=0.69\linewidth]{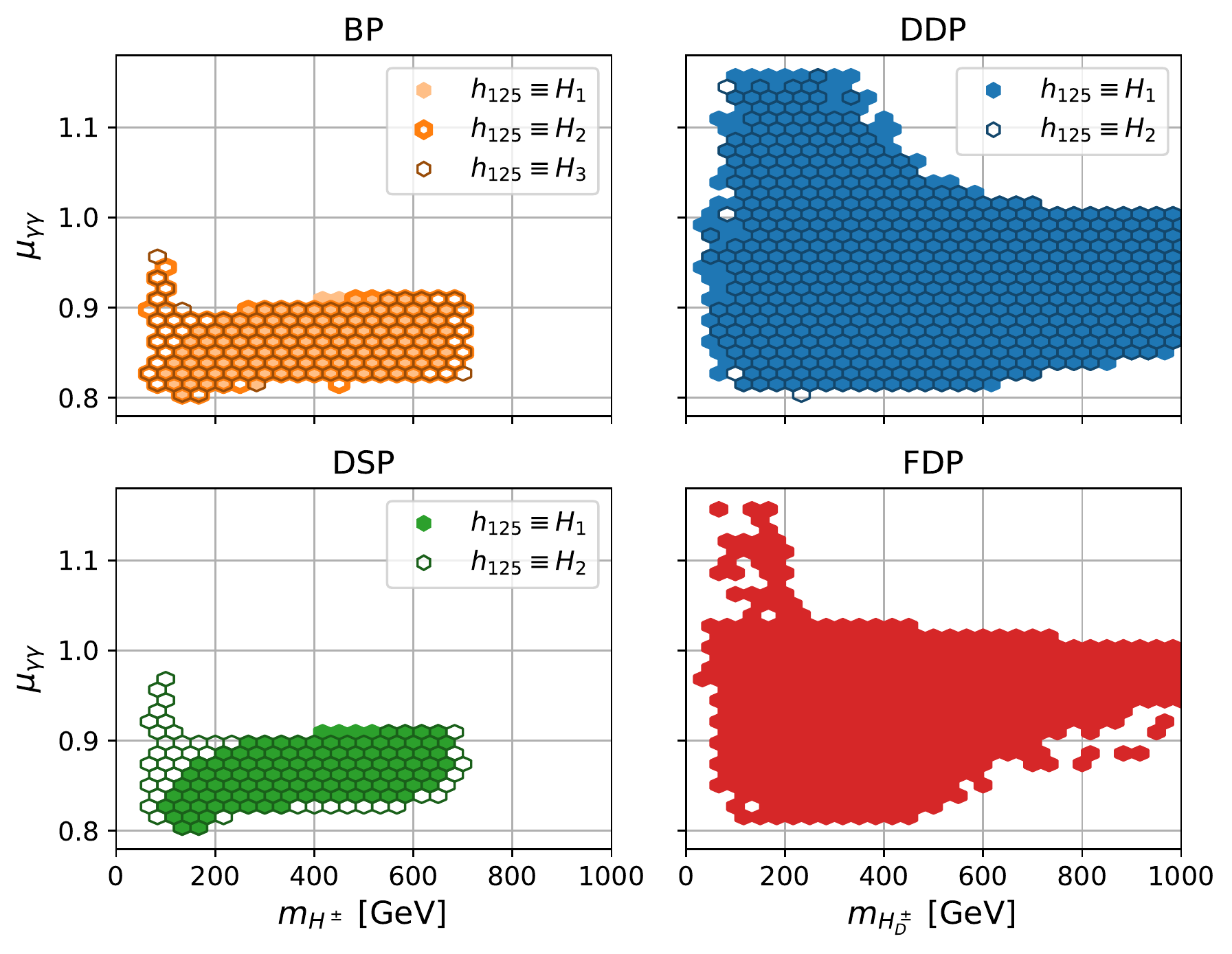}
  \caption{$\mu_{\gamma \gamma}$ as a function of the charged Higgs mass for the four N2HDM phases. %
    }\label{fig:mugaga}
\end{figure}
%%%%%%%%%%%%%%%%%%%%%%%%%%%%%%%%%%%%%%%%%%%%%%%%%%%%%%%
%
Let us start with the 125 GeV Higgs coupling measurements. All phases
have an alignment limit, that is, there is a set of values for which
the $h_{125}$ couplings to fermions and gauge
bosons are exactly the SM ones. Hence, in order to be
  able to distinguish between the phases we need a decay with a new
contribution from a coupling which does not exist the SM and
originates from the Higgs potential. Such is the case of
the $h_{125} \to \gamma \gamma$
decay, which has a contribution from the $h_{125} H^+ H^-$ vertex. In
Fig.~\ref{fig:mugaga} we present $\mu_{\gamma \gamma}$ as a function
of the charged Higgs mass for the four N2HDM phases.
In the BP and DSP phases, which are the ones with charged scalars in the visible sector, the value of $\mu_{\gamma \gamma}$ is always below 0.98 and for charged Higgs masses above 150 GeV
the value is about 0.9 or below. The reason for the low values of $\mu_{\gamma \gamma}$ is due to setting $m_{12}^2 = 0$ (this is the soft breaking term that is usually
included in the broken phase of the 2HDM and in that of the N2HDM). In this limit, the contribution from the $h_{125} H^+ H^-$ vertex, close
to the alignment limit, is always negative, reducing
the diphoton branching ratio of $h_{125}$ relative to its SM value. In the  DDP and FDP the same vertex is proportional to the free parameter $m_{22}^2$,
allowing for both negative and positive contributions.
 Therefore, the freedom in the coupling is lost due to $m_{12}^2 =0$ in the visible phases, while in the dark phases the
free mass parameter leads to a weaker constraint.

The presently measured value of $\kappa_\gamma = \sqrt{\Gamma ({h_{\text{NEW}} \to \gamma \gamma})/\Gamma ({h_\text{SM} \to \gamma \gamma})} $ is $0.97 \pm 0.07$ (at $1\sigma$)\cite{deBlas:2018tjm} while
the HL-LHC 68\% probability sensitivity to the same coupling modifier ranges from $\pm 0.023$ to $\pm 0.016$~\cite{Cepeda:2019klc}.
 This means that if by the end of the LHC high luminosity run the central  value of the branching ratio of the Higgs boson to two photons is
 very close to the SM value and taking into account the predicted errors it is likely
 that the BP and the DSP will be excluded. The only possible exception is the light charged Higgs region
 which on the other hand will also be much more constrained by the end of the high luminosity phase
  by direct searches for charged Higgs bosons.

\subsection{Search for new scalars}
%
%%%%%%%%%%%%%%%%%%%%%%%%%%%%%%%%%%%%%%%%%%%%%%%%%%%%%%%
\begin{figure}
  \centering
\includegraphics[width=0.49\linewidth]{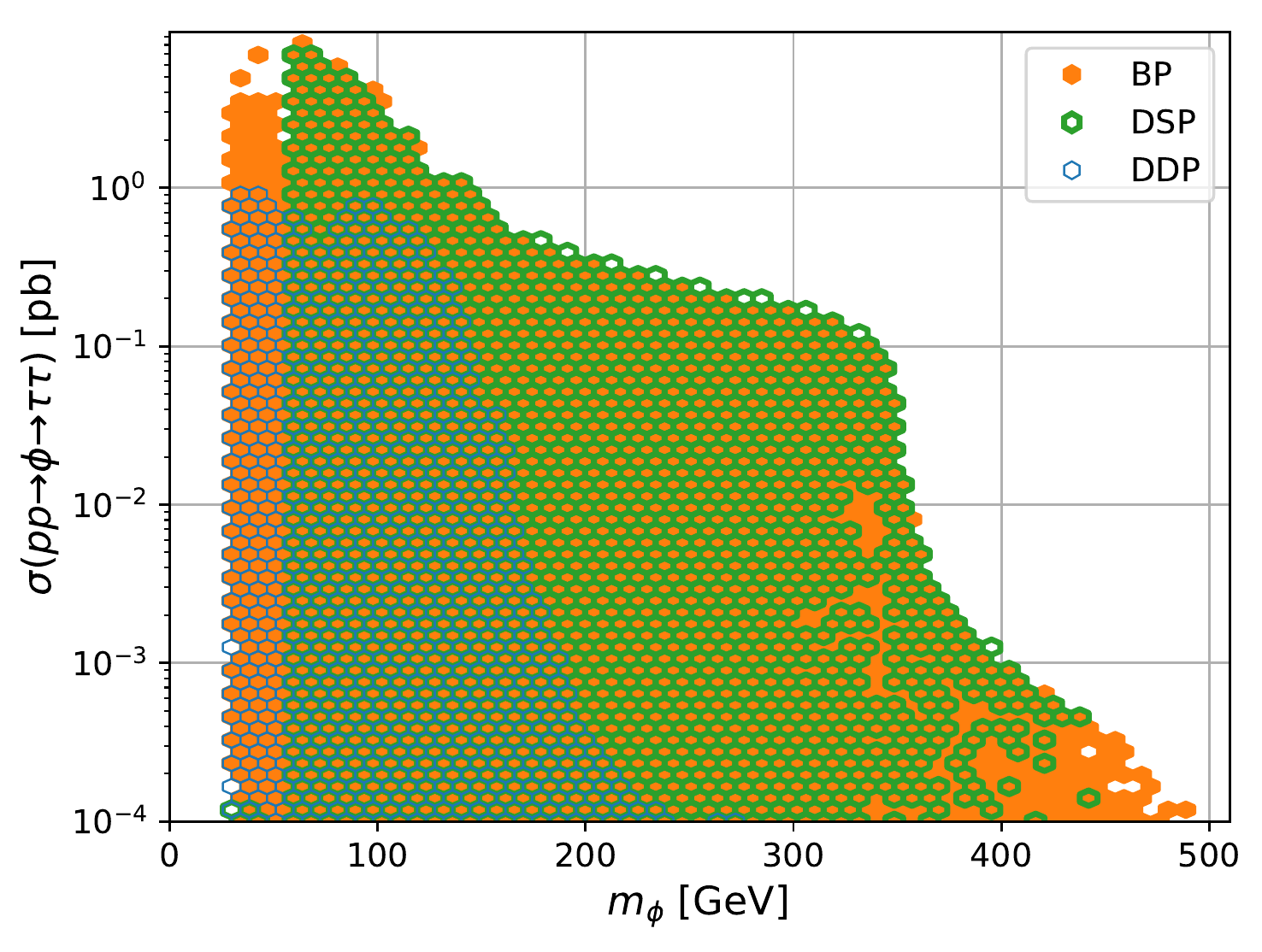}
\includegraphics[width=0.49\linewidth]{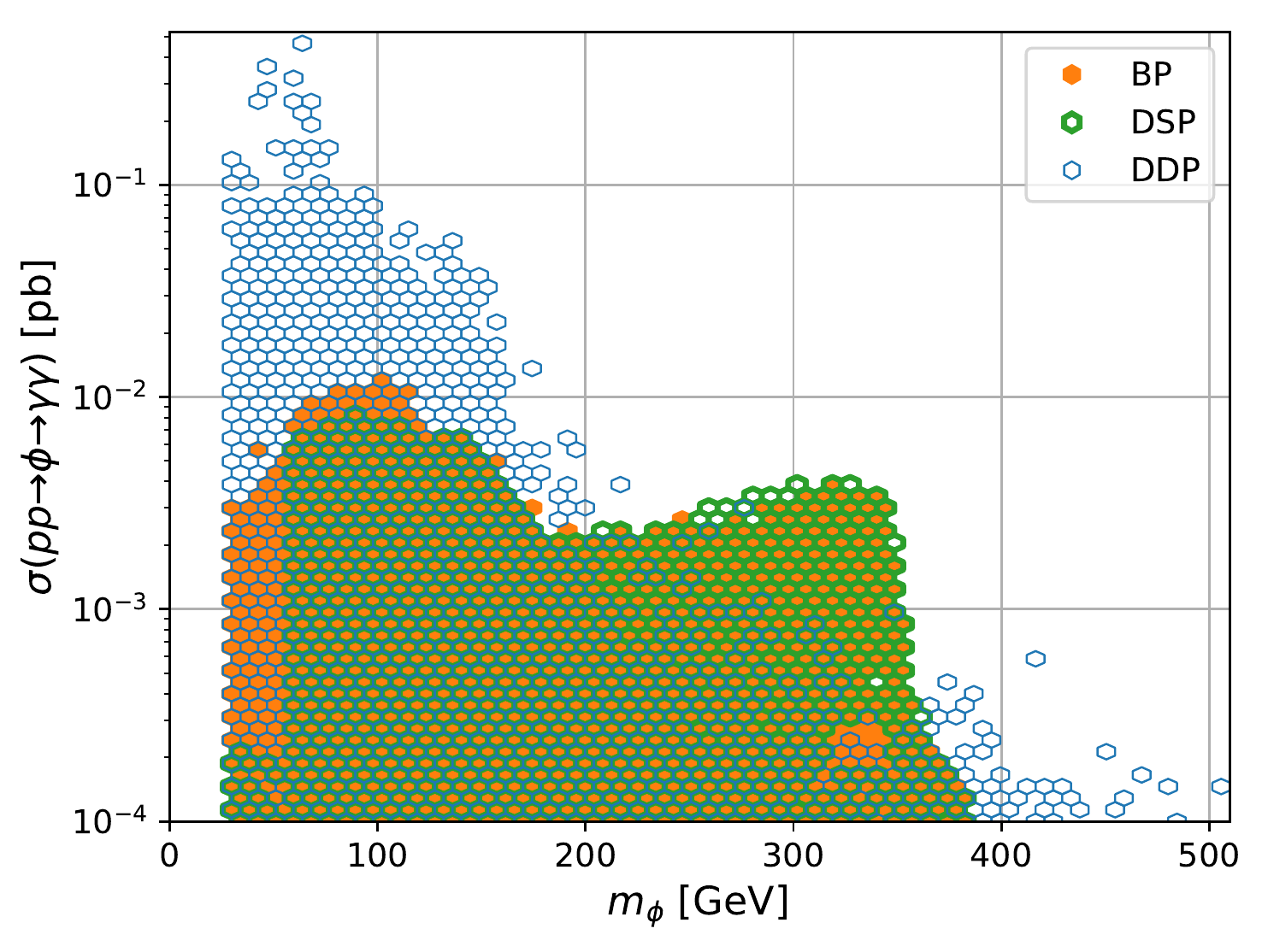}
  \caption{Production cross section for any of the the non-SM like
    Higgs with subsequent decay to $\tau^+ \tau^-$ (left) and $\gamma
    \gamma$ (right) as a function of the scalar
    mass, for the BP, DSP and DDP. $\phi$ stands for any of the CP-even scalars in each phase, other than the 125 GeV one.
    }\label{fig:newsca}
\end{figure}
%%%%%%%%%%%%%%%%%%%%%%%%%%%%%%%%%%%%%%%%%%%%%%%%%%%%%%%
%
As previously discussed, there are some particularities that are
specific to each model. The FDP can only be distinguished from the SM
through the amount of missing energy in collider dark
 matter searches because it contains no new
particles in the visible sector. Charged Higgs bosons in the visible
sector are only possible in the BP
and in the DSP\@. In order to distinguish these two phases one would need to look again into the amount of missing energy in searches for dark matter events at colliders. A feature that all of the phases except
the FDP have in common is the existence of at least one additional, visible neutral scalar.

In Fig.~\ref{fig:newsca} we show the production cross section for the
non-SM like neutral Higgs with subsequent decay to $\tau^+ \tau^-$ (left) and $\gamma \gamma$ (right).
In the phases where we have more than one visible scalar, we take all
possibilities into account, that is, all CP-even scalars are considered.
The decay to $\tau^+ \tau^-$ is chosen because it represents the general
behaviour of the decays to fermions and the $b \bar b$ final state is much harder to resolve due to the background. The most relevant features of fermion final states are
as follows. Below $m_{h_{125}}/2$ the BP accommodates the largest possible rates because decays of the Higgs to dark matter are not possible in this phase. Still, in the DDP values of the cross section as large
as 1 pb are still possible. On the other hand the DDP has less freedom in the visible sector and therefore cross sections for masses above about 230 GeV are
already below 0.1 fb. Above $m_\phi/2$ the BP and DSP are almost indistinguishable because their visible sectors are very similar to a 2HDM, a feature that is reinforced by the tight constraints on the
$h_{125}$ couplings and existing constraints from Higgs searches.

On the right plot of Fig.~\ref{fig:newsca} we can see the decays to $\gamma
\gamma$. In this case the DDP allows for substantially larger cross sections
than the other phases that can even go up to 1 pb for masses below 100 GeV.
Note that although the DDP has less freedom in the visible sector it has more
freedom in the dark sector and this is reflected in the couplings of the dark
charged Higgs boson to the visible scalars. This can not only lead to the
previously discussed large effects in $\mu_{\gamma\gamma}$ for $h_{125}$ but can
also significantly enhance the $pp\to\phi\to\gamma\gamma$ cross sections shown here.
If such a signal is seen with rates above $10^{-2}$ pb all phases except for the
DDP would be excluded.

%
%%%%%%%%%%%%%%%%%%%%%%%%%%%%%%%%%%%%%%%%%%%%%%%%%%%%%%%
\begin{figure}
  \centering
\includegraphics[width=0.49\linewidth]{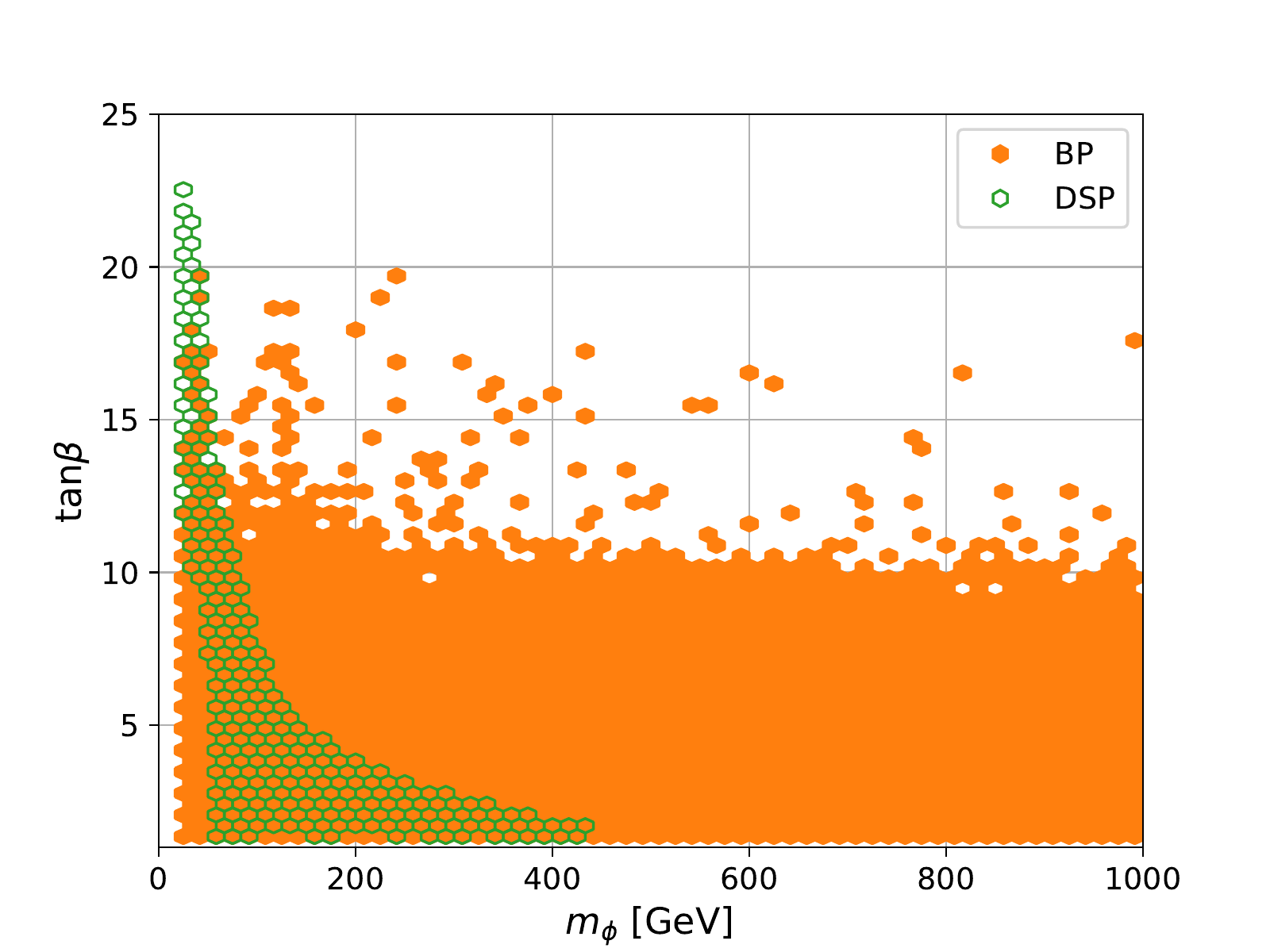}
\includegraphics[width=0.49\linewidth]{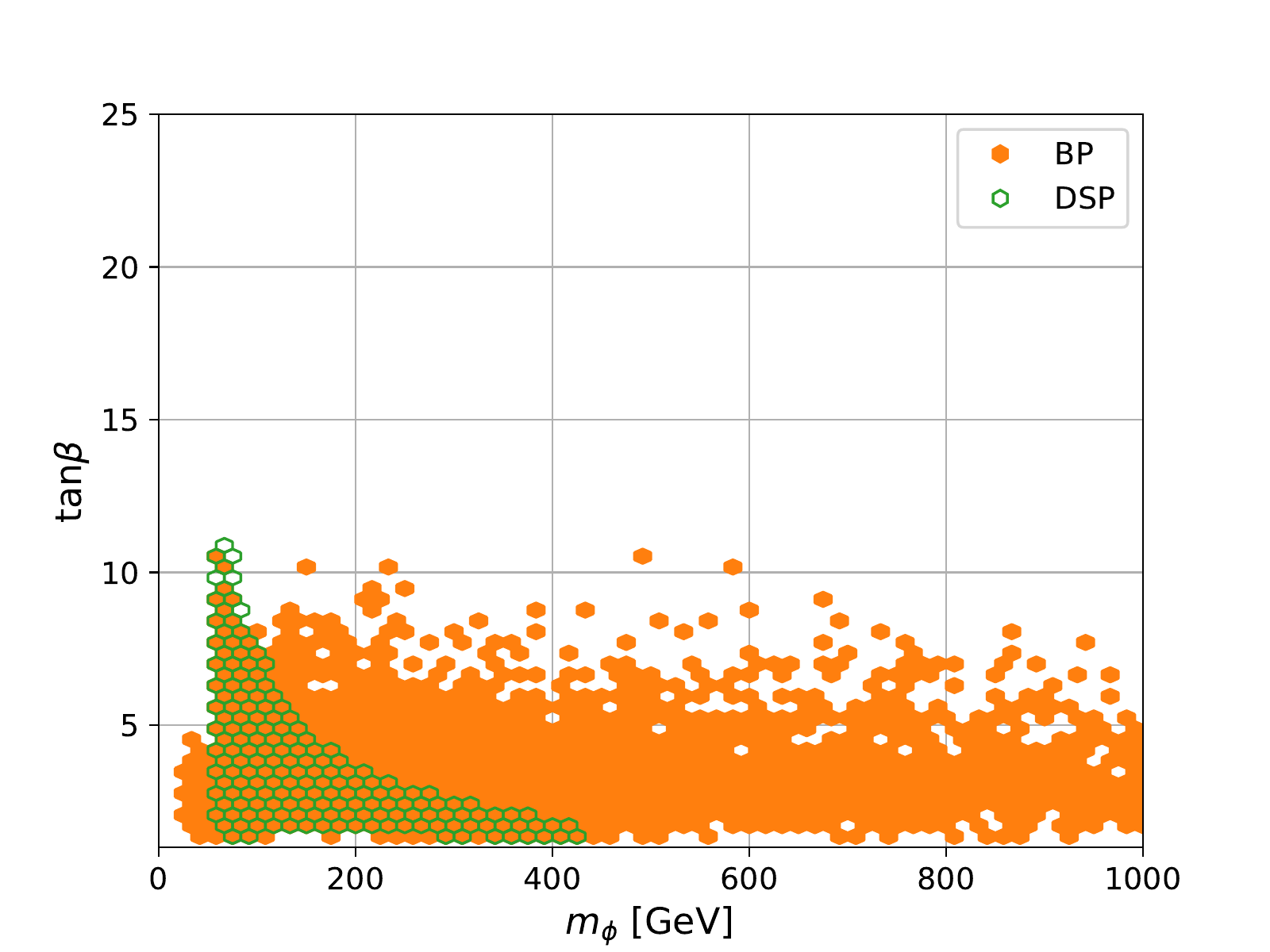}
  \caption{$\tan \beta$ as a function of the mass of any of the
    neutral non-SM like Higgs bosons with the present constraints
    (left) and taking the value of $c (h_{125} VV)$ to be  1  within
    $10^{-3}$ (right), for the BP and DSP. Again, $\phi$ stands for any of the CP-even
    scalars in each phase, other than the 125 GeV one.
    }\label{fig:tanbeta}
\end{figure}
%%%%%%%%%%%%%%%%%%%%%%%%%%%%%%%%%%%%%%%%%%%%%%%%%%%%%%%
%

Let us finally comment on the behaviour of the model very close to the
alignment limit.  As shown in Ref.~\cite{Gorczyca:2011he}, the 2HDM with an exact
$\mathbb{Z}^{(1)}_2$ symmetry, and in the alignment limit where $\sin(\beta - \alpha) = 1$
(or more generally $c (h_{125} VV)=1$), always has a value of
$\tan \beta\lesssim 6$. In that reference they conclude that the limit
arises from a combination of theoretical constraints together with taking the
alignment limit. The left panel of  Fig.~\ref{fig:tanbeta} shows, for the BP and
the DSP, $\tan \beta$ as a function of the mass of any of the CP-even, neutral scalars
other than $h_{125}$, with all present experimental and theoretical
constraints taken into account (note that there is no $\tan \beta$ in the DDP).
The right panel is the same plot as the one on the left with the extra constraint of forcing $c (h_{125} VV)$ to be within $10^{-3}$ of the value 1. Hence, although we have more freedom in our model because
we have an extra singlet field, Fig.~\ref{fig:tanbeta} shows that the allowed value of $\tan \beta$ is reduced as we approach the alignment limit. This has important consequences to corner the model using
all experimental data. As an example, the experimental searches for charged Higgs bosons include the vertex $tbH^\pm$ which, in Yukawa sectors of Type I, is always
proportional to $1/\tan \beta$. Therefore, it will be very hard to access even the light charged Higgs for very large values of $\tan \beta$. However, with the
restriction from the right plot of Fig.~\ref{fig:tanbeta}, moving close to alignment reduces the allowed value of $\tan \beta$. Hence, if $\tan \beta$ is not too large
it is more likely that the charged Higgs production cross section will be within experimental reach.
The more constraints we can find from other sources the closer we will be to exclude a given phase.

\section{Conclusions}
 \label{sec:conclusions}

In this work we  have studied the four phases of the N2HDM --- three of which have dark matter candidates.
For the phases to be comparable, we have considered Yukawa sectors of type I and set $m_{12}^2=0$. The absence of this term makes the scalar potential correspond to an Inert Doublet Model extended by a real singlet field.
% In fact this term is also not allowed in
%the N2HDM version of the Inert Doublet Model where the singlet field is absent.
The different phases have
the same scalar potential and degrees of freedom but the fields in the dark sector vary from the FDP where all the extra degrees of
freedom are in dark sector to the BP which has no dark matter candidate. The analysis of the vacuum structure
of the four phases has shown an interesting behaviour of the possible neutral minima. We have shown that
if a minimum  in the BP or FDP exists, it is the global minimum of the theory. In that case all other
stationary points of different phases lie above it and are saddle points.
However, the same is not true for minima in the DDP and DSP - they can coexist in the potential, and neither is guaranteed
to be deeper than the other.

Our main goal was to understand if the different phases could be probed and distinguished
by combining the available experimental data and the one from future searches at colliders
with that from dark matter experiments. We have generated samples of points for each phase which
take into account the most up-to-date experimental data and also all relevant theoretical constraints.
From the dark matter point of view, and in particular the direct detection bounds, all phases have valid points all the
way to the neutrino floor. Hence, future direct detection experiments will not play a major role in constraining
the parameter space of the model. As for dark matter relic density, all except the Dark Doublet Phase, have candidates
for dark matter that saturate the relic density for a large range of dark masses. The DDP behaves very much like the Inert Doublet Model where, as
previously discussed, the relic density cannot be saturated for dark matter masses between about 100 GeV and 500 GeV.
We have then looked for the effect of the Higgs coupling measurements and for the search for new particles
at the LHC\@. Our main conclusions on what can we learn from the LHC are as follows:
\begin{itemize}

\item Finding a charged Higgs would single out the BP and DSP, while the
discovery of any new neutral scalar would exclude the FDP.

\item Visible and dark sector charged Higgs bosons have very different
impacts on the decays of the neutral scalars into $\gamma\gamma$. Visible
$H^\pm$ always suppress $\mu_{\gamma\gamma}$ compared to the SM, while dark
$H^\pm_D$ have more freedom in their couplings and could enhance or suppress the rate. As a
result, a measurement of $\mu_{\gamma \gamma}$ at the end of the HL-LHC or future
collider could very well exclude the BP and the DSP.

\item In case a new scalar is found there are regions of parameter space where
the 3 phases, BP, DDP and DSP could be distinguished in the decay to $\tau^+ \tau^-$. Due to
the dark charged Higgs, the DDP can predict very large rates for a new scalar
decaying into $\gamma\gamma$ and may be probed there.
\item
If nothing is discovered and the 125 GeV Higgs couplings are very close to
the SM values, the FDP will always remain a possibility.

\end{itemize}

\appendix

%% ==============================
\section{Dark Doublet Phase}
\label{app:DDP}
%% ==============================
In this section, we present for the DDP the relation between the Lagrangian
parameters and the physical parameters. First, the physical masses can be
written as
\begin{subequations}
\begin{align}
\label{eq:ID_mH1sq}
m^2_{H_{1}} %&= \mathcal{R}_{13} v_s (-\mathcal{R}_{22} v_s \lambda_6 + \mathcal{R}_{23} v \lambda_8) + \mathcal{R}_{12} v (\mathcal{R}_{23} v \lambda_2 - \mathcal{R}_{22} v_s \lambda_8)\\
&=  v^2 \cos^2\alpha\, \lambda_1 + v_s^2 \sin^2\alpha\, \lambda_6 + 2v v_s \sin\alpha\,\cos\alpha\, \lambda_7\,,\\
\label{eq:ID_mH2sq}
m^2_{H_{2}} %&= \mathcal{R}(1,2) v_s (\mathcal{R}(2,3) v_s \lambda_6 + \mathcal{R}(2,2) v \lambda_8) - \mathcal{R}(1,3) v (\mathcal{R}(2,2) v \lambda_2 + \mathcal{R}(2,3) v_s \lambda_8)\\
 &= v^2 \sin^2\alpha\,\lambda_1 + v_s^2 \cos^2\alpha\,\lambda_6 - 2 v v_s \sin\alpha\,\cos\alpha\, \lambda_7\,,\\
\label{eq:ID_mHDsq}
m^2_{H_{D}} &= \dfrac{1}{2} (2 m_{22}^2 + v^2 (\lambda_3 + \lambda_4 + \lambda_5) + v_s^2 \lambda_8)\,,\\
\label{eq:ID_mADsq}
m^2_{A_{D}}
&= \dfrac{1}{2} (2 m_{22}^2 + v^2 (\lambda_3 + \lambda_4 - \lambda_5) + v_s^2 \lambda_8)\,,\\
\label{eq:ID_mHcDsq}
m^2_{H_{D}^{\pm}}
&= \dfrac{1}{2} (2 m_{22}^2 + v^2 \lambda_3 + v_s^2 \lambda_8)\,,
\end{align}
\end{subequations}
which leads to the following relations between the parameters
\begin{subequations}
\begin{align}
\lambda_1 =& \frac{1}{v^2}\left(\sum_i m_{H_{i}}^2 \mathcal{R}^2_{i1}\right),\\
\lambda_3 =& \frac{1}{v^2}\left(2\left(m_{H^{\pm}_D}^2 - m_{22}^2\right) - v_s^2\, \lambda_8\right), \\
\lambda_4 =& \frac{1}{v^2}\left(m_{A_D}^2 + m_{H_{D}}^2 - 2 m_{H^{\pm}}^2\right),\\
\lambda_5 =& \frac{1}{v^2}\left(m_{H_{D}}^{2}-m_{A_D}^{2}\right),\\
\lambda_6 =& \frac{1}{v_s^2}\left(\sum_i m_{H_{i}}^2 \mathcal{R}^2_{i3}\right),\\
\lambda_7 =& \frac{1}{v v_s}\left(\sum_i m_{H_{i}}^2 \mathcal{R}_{i1} \mathcal{R}_{i3}\right),
\end{align}
\end{subequations}
where $\mathcal{R}_{ij}$ is the $i,j$ element of the mixing matrix in \eqref{eq:inertdoubletscalarmixingmatrix}.
The parameters $m_{22}^2$, $\lambda_2$ and $\lambda_8$ cannot be expressed through physical parameters and thus remain independent parameters in the physical parameter set of the DDP.

\subsection{Triple-Higgs Couplings}
\label{app:IDPTripleHiggsCouplings}
The triple-Higgs couplings $g(X_i X_j X_k)$ in the DDP are defined as,
\begin{align}
g(X_i X_j X_k) = \frac{\partial^3\mathcal{L}}{\partial X_i \partial X_j \partial X_k}\,,\label{eq:def_coupling}
\end{align}
with $X_{i/j/k} \in \left\{H_1, H_2, H_D, A_D, H^\pm_D\right\}$.
All couplings with an odd number of dark Higgs bosons vanish due to the conserved dark parity. The non-zero triple-Higgs couplings are the following, where the indices $i,j$ can only be $\left\{1,2\right\}$
and denote the visible CP-even Higgs bosons $H_1$ or $H_2$, respectively:
\begin{align}
\begin{split}
g(H_i H_i H_i) &= 3 \,\lambda_1 \, v  \mathcal{R}^3_{i1}
				  + 3 \,\lambda_6 \, v_s  \mathcal{R}^3_{i3}\\
			    &\quad + 3 \lambda_7 \left( v \mathcal{R}_{i1}  \mathcal{R}^2_{i3} + v_s  \mathcal{R}_{i3}  \mathcal{R}^2_{i1}\right),
\end{split}\\
\begin{split}
g(H_i H_j H_j) &= 3 \,\lambda_1 \, v  \mathcal{R}_{i1}  \mathcal{R}^2_{j1}
				  + 3 \,\lambda_6 \, v_s  \mathcal{R}_{i3}  \mathcal{R}^2_{j3} \\
			    &\quad + \lambda_7 \left[v  \left( \mathcal{R}_{i1}  \mathcal{R}^2_{j3} + 2 \mathcal{R}_{i3}  \mathcal{R}_{j1}  \mathcal{R}_{j3}\right)\right.\\
			    &\quad  \left.\quad\quad\,\, + v_s  \left(\mathcal{R}_{i3}  \mathcal{R}^2_{j1} + 2 \mathcal{R}_{i1}  \mathcal{R}_{j1}  \mathcal{R}_{j3}\right)\right],
\end{split}\\
g(H_i H_D H_D) &= \dfrac{2}{v}\, \left(m_{H_{D}}^{2}-m_{22}^{2}\right) \mathcal{R}_{i1} + \lambda_8 \, \dfrac{v_s}{v} \,\left(v   \mathcal{R}_{i3} - v_s  \mathcal{R}_{i1}\right),\\
g(H_i H_D^{+} H_D^{-}) &= \dfrac{2}{v}\,\left(m_{H^{\pm}_D}^{2}-m_{22}^{2}\right) \mathcal{R}_{i1} + \lambda_8\, \dfrac{v_s}{v}\,\left(v \mathcal{R}_{i3}-v_s \mathcal{R}_{i1}\right),\\
g(H_i A_D A_D) &= \dfrac{2}{v}\,\left(m_{A_D}^{2}-m_{22}^{2}\right) \mathcal{R}_{i1} + \lambda_8\, \dfrac{v_s}{v}\,\left(v \mathcal{R}_{i3}-v_s \mathcal{R}_{i1}\right).
\end{align}

%% ==============================
\section{Dark Singlet Phase}
\label{app:DSP}
%% ==============================
In this section, we present for the DSP the formulae for the masses
and the relation between
the gauge basis and the physical basis. The expressions for the masses are
\begin{subequations}
\begin{align}
\label{eq:DS_mH1sq}
m^2_{H_{1}} &= \frac{m_{12}^2}{v_1 v_2} (v_1 \cos\alpha + v_2 \sin\alpha)^2 \\
&\quad + \lambda_1 v_1^2 \cos^2\alpha \notag
+ \lambda_2 v_2^2 \sin^2\alpha
- 2\lambda_{345} v_1 v_2 \cos\alpha \sin\alpha\,,\\
% =& +m_{12}^2 \left(\frac{\cos\alpha}{vs_\beta} + \frac{\sin\alpha}{vc_\beta}\right)\\
% &+
% & -\lambda_1 v_1^2 \mathcal{R}(1,1) \mathcal{R}(2,2)
% 	+\lambda_2 v_2^2 \mathcal{R}(1,2) \mathcal{R}(2,1) \\
% 	&+ \lambda_{345} v_1 v_2 (\mathcal{R}(1,1) \mathcal{R}(2,1) - \mathcal{R}(1,2) \mathcal{R}(2,2))\\
\label{eq:DS_mH2sq}
m^2_{H_{2}} &=\frac{m_{12}^2}{v_1 v_2} (v_1 \sin\alpha - v_2 \cos\alpha)^2 \\
&\quad+ \lambda_1 v_1^2 \cos^2\alpha \notag
+ \lambda_2 v_2^2 \sin^2\alpha
+ 2\lambda_{345} v_1 v_2 \cos\alpha\sin\alpha\,,\\
% 	=& \lambda_1 v_1^2 \mathcal{R}(1,2) \mathcal{R}(2,1)
% 	- \lambda_2 v_2^2 \mathcal{R}(1,1) \mathcal{R}(2,2)\\
% 	&+ \lambda_{345} v_1 v_2 (\mathcal{R}(1,2) \mathcal{R}(2,2) - \mathcal{R}(1,1) \mathcal{R}(1,2))\\
\label{eq:DS_mHDsq}
m^2_{H_{D}} &= \dfrac{1}{2} (2 m_s^2 + v_1^2 \lambda_7 + v_2^2 \lambda_8)\,,\\
\label{eq:DS_mAsq}
m^2_{A} &= -v^2 \lambda_5 + \frac{m_{12}^2}{s_\beta c_\beta}\,,\\
\label{eq:DS_mHcsq}
m^2_{H^{\pm}} &= - \dfrac{1}{2} v^2 (\lambda_4 + \lambda_5) + \frac{m_{12}^2}{s_\beta c_\beta}\,.
\end{align}
\end{subequations}

The relations between the two sets of parameters are
\begin{subequations}
\begin{align}
m_{S}^2 &=  -\dfrac{1}{2}\left(v_1^2 \lambda_7 + v_2^2 \lambda_8 - 2 m_{H_D}\right),\\
\lambda_1 &= \frac{1}{v^2 c^2_\beta}\left[\left(\sum_i m_{H_{i}}^2 \mathcal{R}^2_{i1}\right) -m_{12}^2 \frac{s_\beta}{c_\beta}\right],\\
\lambda_2 &= \frac{1}{v^2 s^2_\beta}\left[\left(\sum_i m_{H_{i}}^2 \mathcal{R}^2_{i2}\right) -m_{12}^2 \frac{c_\beta}{s_\beta}\right],\\
\lambda_3 &= \frac{1}{v^2 c_\beta s_\beta}\left[\left(\sum_i m_{H_{i}}^2 \mathcal{R}_{i1} \mathcal{R}_{i2}\right) - m_{12}^2\right] + \frac{2}{v^2}\, m_{H^\pm}^2\,,\\
\lambda_4 &= \frac{1}{v^2}\left(m_A^2 - 2 m_{H^\pm}^2\right) + \frac{1}{v^2 c_\beta s_\beta}\, m_{12}^2\,,\\
\lambda_5 &= - \frac{1}{v^2}\, m_A^2 + \frac{1}{v^2 c_\beta s_\beta}\,m_{12}^2\,,
\end{align}
\end{subequations}
where $\mathcal{R}_{ij}$ is the $i,j$ element of the mixing matrix in \eqref{eq:DSPmixingmatrix}.
The parameters $\lambda_6$, $\lambda_7$ and $\lambda_8$ cannot be expressed through physical parameters and thus remain independent parameters in the physical parameter set of the DSP.

\subsection{Triple-Higgs Couplings}
\label{app:DSPTripleHiggsCouplings}
We now present the triple-Higgs couplings $g(X_i X_j X_k)$ in the DSP.
The definition of the coupling $g(X_i X_j X_k)$ is given in \eqref{eq:def_coupling} with $X_{i/j/k} \in \{H_1, H_2, H_D,\allowbreak A, H^\pm\}$.
All couplings with an odd number of $H_D$ vanish due to the conserved dark parity.
The non-zero triple-Higgs couplings --- with $i,j$ again reserved for the visible sector Higgs
bosons --- are
\begin{align}
g(H_iH_iH_i) &= 3v \left[
c_\beta \left(
\mathcal{R}^3_{i1} \lambda_1 + \mathcal{R}_{i1}\mathcal{R}^2_{i2} \lambda_{345} \right)
+ s_\beta \left(
\mathcal{R}^3_{i2} \lambda_2 + \mathcal{R}_{i2}\mathcal{R}^2_{i1} \lambda_{345}
\right)
\right]\,,\\
\begin{split}
g(H_iH_jH_j) &=
v \Big[c_\beta \left(
3 \mathcal{R}_{i1} \mathcal{R}^2_{j1} \lambda_1
+ (3 \mathcal{R}_{i2}  \mathcal{R}_{j1} \mathcal{R}_{j2} + \mathcal{R}_{i1}) \lambda_{345}
\right)\\
&\qquad+ v s_\beta \left(
3 \mathcal{R}_{i2} \mathcal{R}^2_{j2} \lambda_2
+ (3 \mathcal{R}_{i1} \mathcal{R}_{j1} \mathcal{R}_{j2} + \mathcal{R}_{i2}) \lambda_{345}
\right)\Big]\,,
\end{split}\\
\begin{split}
g(H_iAA) &= v \Big[
c_\beta\left(
c_\beta s_\beta \mathcal{R}_{i2} \left(\lambda_2 -2\lambda_5\right)
+ c^2_\beta \mathcal{R}_{i1} \lambda_{34-5}
\right)\\
&\qquad+ s_\beta \left(
c_\beta s_\beta \mathcal{R}_{i2} \left(\lambda_1 -2 \lambda_5\right)
+ s^2_\beta \mathcal{R}_{i2} \lambda_{34-5}
\right)
\Big]\,,
\end{split}\\
\begin{split}
g(H_iH^+H^-) &= v \Big[
c_{\beta} \left(
s^2_\beta \mathcal{R}_{i1} \lambda_1
+ c^2_\beta \mathcal{R}_{i1} \lambda_3
- c_{\beta} s_{\beta} \mathcal{R}_{i2} \left(\lambda_4 + \lambda_5\right)
\right)\\
&\qquad + s_\beta \left(
 c^2_\beta \mathcal{R}_{i2} \lambda_2
+ s^2_\beta \mathcal{R}_{i2} \lambda_3
- c_{\beta} s_{\beta} \mathcal{R}_{i1} \left(\lambda_4 + \lambda_5\right)
\right)
\Big]\,,
\end{split}\\
g(H_iH_DH_D) &= v\left[c_\beta \mathcal{R}_{i,1} \lambda_7 + s_\beta \mathcal{R}_{i,2} \lambda_8\right]\,.
\end{align}

%%%%%%%%%%%%%%%%%%%%%%%%%%%%%%%%%%%%%%%%%%%%%%%%%%%%%%%
\subsubsection*{Acknowledgments}

We acknowledge discussions with Igor Ivanov, Tania Robens and Dorota Sokolowska.
PF and RS are supported by the Portuguese Foundation for Science and Technology (FCT), Contracts UIDB/00618/2020,
UIDP/00618/2020, PTDC/FIS-PAR/31000/2017 and CERN/FIS-PAR/0002/2017, and by the HARMONIA project, contract UMO-2015/18/M/ST2/00518.
JW has been funded by the European Research Council (ERC)
under the European Union's Horizon 2020 research and innovation
programme, grant agreement No 668679. MM is supported by the
BMBF-Project 05H18VKCC1.

\vspace*{0.5cm}

%%%%%%%%%%%%%%%%%%%%%%%%%%%%%%%%%%%%%%%%%%%%%%%%%%%%%%%%%%%%
\vspace*{1cm}
\bibliographystyle{h-physrev}
%\bibliography{TIPl}
\bibliography{TIPv1.bib}

\end{document}